\documentclass[twocolumn]{aa}
\usepackage[dvips]{color}
\usepackage{graphicx}
\usepackage{lscape}
\usepackage{txfonts}
\usepackage{natbib}
\usepackage{ulem}

\bibpunct{(}{)}{;}{a}{}{,} 

\def\ha{{H$\alpha$}\/} 
\def\lya{{Ly$\alpha$}\/}
\def\hb{{H$\beta$}\/} 
\def\nii{[{\sc N\,ii}]\/} 
\def\oii{{[{{\sc O\,ii}}]\/}} 
 
\def\Oii{{[{{\sc O}\,{\sc ii}}]~$\lambda\lambda$3726,3728}\/} 
\def\Oiii{{[{{\sc O}\,{\sc iii}}]~$\lambda\lambda$4959,5007}\/}

\def\Hii{{{\sc H\,\sc ii}}\/}

\def\na{{$N_{810}$}\/}
\def\nb{{$N_{817}$}\/}
\def\nc{{$N_{824}$}\/}
\def\med{{$M_{815}$}\/}
\def\mminn{\mbox{\med--$N$}}

\def\simlt{\lower.5ex\hbox{$\; \buildrel < \over \sim \;$}}
\def\simgt{\lower.5ex\hbox{$\; \buildrel > \over \sim \;$}}
\def\sext{{\sl SExtractor}}
\def\taurus2{{\sc Taurus-2}}
\def\lineunits{erg\,s$^{-1}$\,cm$^{-2}$}

\def\ergsPerAng{erg\,s$^{-1}$\,cm$^{-2}$\,\AA$^{-1}$}
\def\ergs2band{erg\,s$^{-1}$\,cm$^{-2}$\,band$^{-1}$}
\def\ergs{erg\,s$^{-1}$}
\def\lumDens{erg\,s$^{-1}$\,Mpc$^{-3}$}
\def\uJy{$\mu$Jy}
\def\as{$''$}
\def\am{$'$}

\def\d{$^{\circ}$}
\def\MpcPer3{Mpc$^{-3}$}
\def\perMpc3{Mpc$^{-3}$}
\def\Mpc3{Mpc$^{3}$}

\def\kmsMpc{km\,s$^{-1}$\,Mpc$^{-1}$}

\def\kms{km\,s$^{-1}$}

\def\h{$^{\rm h}$}
\def\m{$^{\rm m}$}

\begin{document}
\newcommand{\pow}[2][10]{#1^{#2}}

\title{The Wide Field Imager Lyman-Alpha Search (WFILAS) for Galaxies
  at Redshift $\sim$\,5.7
  \thanks{ Based on observations made with ESO Telescopes at the La
    Silla Observatory (Programmes 67.A-0063, 68.A-0363 and
    69.A-0314).}
}

\subtitle{II. Survey Design and Sample Analysis}

\author{E. Westra\inst{1} \and D.\ Heath Jones\inst{1,2} \and C.~E.\
  Lidman\inst{3} \and K. Meisenheimer\inst{4} \and\\
  R.~M. Athreya\inst{5} \and C. Wolf\inst{6} \and T. Szeifert\inst{3}
  and E. Pompei\inst{3} \and L. Vanzi\inst{3} }

\institute{Research School of Astronomy \& Astrophysics, The
  Australian National University, Cotter Road, Weston Creek ACT 2611,
  Australia, \email{westra@mso.anu.edu.au, heath@mso.anu.edu.au}
  \and Anglo-Australian Observatory, PO Box 296, Epping NSW 1710, Australia
  \and European Southern Observatory, Casilla 19001, Santiago 19,
  Chile, \email{clidman@eso.org, tszeifer@eso.org, epompei@eso.org,
    lvanzi@eso.org}
  \and {Max Planck Institut f\"ur Astronomie, K\"onigstuhl 17,
    69117 Heidelberg, Germany, \email{meise@mpia.de}}
  \and {National Centre for Radio Astrophysics, Tata Institute of
    Fundamental Research Pune University Campus, Post Bag 3,
    Ganeshkhind Pune 411007, India, \email{rathreya@ncra.tifr.res.in}}
  \and {Department of Astrophysics, Denys Wilkinson Building,
    University of Oxford, Keble Road, Oxford, OX1 3RH, U.K.,
    \email{cwolf@astro.ox.ac.uk}}
}

\date{Received 20 January 2006 / Accepted 28 April 2006}

\abstract{Wide-field narrowband surveys are an efficient way of
  searching large volumes of high-redshift space for distant galaxies.}
{We describe the Wide Field Imager Lyman-Alpha Search (WFILAS) over
  0.74\,sq.\ degree for bright emission-line galaxies at $z \sim
  5.7$.}
{WFILAS uses deep images taken with the Wide Field Imager (WFI) on the
  ESO/MPI 2.2\,m telescope in three narrowband (70\,\AA), one
  encompassing intermediate band (220\,\AA) and two broadband filters,
  $B$ and $R$. We use the novel technique of an encompassing
  intermediate band filter to exclude false detections. Images taken
  with broadband $B$ and $R$ filters are used to remove low redshift
  galaxies from our sample.}
{We present a sample of seven \lya\ emitting galaxy candidates, two of
  which are spectroscopically confirmed. Compared to other surveys all
  our candidates are bright, the results of this survey complements
  other narrowband surveys at this redshift. Most of our candidates
  are in the regime of bright luminosities, beyond the reach of less
  voluminous surveys. Adding our candidates to those of another survey
  increases the derived luminosity density by $\sim$30\%. We also
  find potential clustering in the Chandra Deep Field South,
  supporting overdensities discovered by other surveys. Based on a
  FORS2/VLT spectrum we additionally present the analysis of the
  second confirmed \lya\ emitting galaxy in our sample. We find that
  it is the brightest \lya\ emitting galaxy
  (1\,$\times\,10^{-16}$\,\lineunits) at this redshift to date and the
  second confirmed candidate of our survey. Both objects exhibit the
  presence of a possible second \lya\ component redward of the line.}
{}


\authorrunning{E. Westra et al.}

\titlerunning{WFILAS for Galaxies at $z \sim$\,5.7. II.}

\maketitle


\section{Introduction}

Detections of both galaxies and QSOs at $z\,\sim\,6$
\citep{Fan02,Becker01,Djorgovski01} indicate that the Universe was
largely reionised at that epoch. The recent three-year {\it WMAP}
results combined with other cosmological surveys suggest an epoch of
reionisation around $z\,\sim\,10$ \citep{Spergel06}, consistent with
both QSO results \citep{Fan02} and the epoch predicted by structure
formation models \citep{Gnedin97,Haiman98}. While the UV contributions
of QSOs and AGN are almost certainly not responsible for reionisation
\citep{Barger03}, faint star forming galaxies need to exist in
extraordinary numbers if they are to be the cause \citep{Yan04}.
However, analyses of the Hubble Ultra Deep Field failed to find
sufficient numbers of faint galaxies to support this idea
\citep{Bunker04,Bouwens05}. Therefore, it is crucial to investigate
what the contribution to the ionising UV flux is from young stellar
populations of star forming galaxies.

Broadly speaking, two classes of star-forming galaxy dominate high
redshift surveys: Lyman Break Galaxies (LBGs) and Lyman-$\alpha$
Emitters (LAEs). LBG surveys, which now number in the thousands of
objects at $z$\,=\,3 to 5, find clumpy source distributions and a
two-point angular correlation function indicative of strong clustering
\citep{Giavalisco01,Foucaud03,Adelberger03,Ouchi04,Hildebrandt05,Allen05}.
LAEs also show evidence for clustering although many of the LAE
surveys target fields surrounding known sources such as
proto-clusters, radio galaxies and QSOs
\citep[e.g.][]{Steidel00,Moeller01,Stiavelli01,Venemans02,Ouchi05}. On
average, LAEs number 1.5\,$\times$\,$\pow{4}$\,deg$^{-2}$ per unit
redshift down to 1.5\,$\times$\,$\pow{-17}$\,\lineunits\ at
$z\,=\,3.4$ and 4.5 \citep{Hu98}. Also, their consistently small size
($\lesssim$0.6\,$h^{-1}$\,kpc) suggests they are subgalactic clumps
residing in the wind-driven outflows of larger unseen hosts
\citep[e.g.][]{BlandHawthorn04}. Such mechanisms provide a
straightforward means of UV photon escape from the host galaxy,
efficiently reionising the surrounding IGM in a way than ordinary LBGs
can not.

The most efficient way to find LAEs is through imaging surveys using a
combination of broad- and narrowband filters. The advent of wide field
cameras has allowed systematic imaging searches that have been carried
out to build up samples of candidate LAEs at high redshifts
\citep[e.g.][]{Rhoads03,Ajiki03,Hu04,Wang05}. The availability of high
throughput spectrographs on 8 to 10\,m-class telescopes has enabled
the spectroscopic confirmation of these galaxies. Such direct imaging
searches typically cover \mbox{10$^2$\,--\,10$^3$} times the volume of
blind long-slit spectroscopic searches \citep[e.g. Table~4
in][]{Santos04}. Furthermore, candidates from narrowband surveys {\it
  always} have an identifiable emission feature that is well separated
from sky lines courtesy of the filter design. This is in contrast to
other methods, including the widely-used ``dropout'' technique
\citep[e.g.][]{Steidel99}.

The narrowband filter design leads to a higher candidate LAE selection
efficiency than other techniques. The only way to secure the
identification of the emission line is spectroscopic follow-up. The
most common low redshift interlopers are the emission line doublets of
\Oii\ and \Oiii. These can be identified by obtaining spectra with a
resolution $R\gtrsim1500$ to separate the line pair. Other emission
lines, such as \ha\ and \hb, can be identified by neighbouring lines.
The narrowband technique has been successfully applied by many authors
in order to discover galaxies at redshift 5$-$6
\citep[e.g.][]{Ajiki03,Maier03,Rhoads03,Dawson04,Hu04} and to locate
galaxies at redshift 6$-$7 \citep{Cuby03,Kodaira03,Stanway04}.
Likewise, we employ the narrowband technique in the Wide Field Imager
Lyman-Alpha Search (WFILAS) to find galaxies at $z\sim5.7$. In Paper I
in this series \citep{Westra05}, we described a compact LAE at
$z=5.721$ discovered by our survey.

In this Paper, we describe the survey design and sample analysis of
WFILAS. In Sect.~\ref{sec:observations} we describe the scope of the
survey and the observing strategy. The data reduction is described in
Sect.~\ref{sec:datared}. Section~\ref{sec:selection} outlines the
candidate selection and Sect.~\ref{sec:candcat} outlines sample
properties and comparison to other surveys. We discuss the
spectroscopic follow-up of two candidates in
Sect.~\ref{sec:confirmed}. Throughout this paper we assume a flat
Universe with $(\Omega_{\rm m}, \Omega_{\Lambda}) = (0.3,0.7)$ and a
Hubble constant $H_0 = 70$\,\kmsMpc. All quoted magnitudes are in the
{\it AB} system \citep{Oke83}\footnote{$m_{AB} = -2.5\,\log\,f_\nu -
  48.590$, where $m_{AB}$ is the {\it AB} magnitude and $f_\nu$ is the flux
  density in ergs\,s$^{-1}$\,cm$^{-2}$\,Hz$^{-1}$}.

\section{WFILAS Survey Design and Observations}
\label{sec:observations}

\begin{table*}[!htbp]
\begin{minipage}[htb]{\textwidth}
\begin{center}
  \renewcommand{\thefootnote}{{\it\alph{footnote}}}
\begin{tabular}{p{5.0cm}cccccc} 
\hline\hline
Survey                 & Fields & Total Area   & Narrowband & Filter       & Co-moving      & Narrowband Detection \\
                       &        & (sq.~degree) & Filters    & Width (\AA)  & Volume (\Mpc3) & Limit (\uJy)  \\
\hline
LALA \citep{Rhoads01}  & 1      & 0.19         & 2          & 75           & 0.2$\times$10$^6$     & 0.41  \\
CADIS \citep{Maier03}  & 4      & 0.11         & 8$-$9\footnotemark[1]           & 20           & 0.04$\times$10$^6$    & 3.33  \\
A03 \citep{Ajiki03}    & 1      & 0.26         & 1         & 120          & 0.2$\times$10$^6$     & 0.14  \\
SSA22 \citep{Hu04}     & 1      & 0.19         & 1          & 120          & 0.2$\times$10$^6$     & 0.30  \\
\\
WFILAS (this paper)    & 3      & 0.74         & 3\footnotemark[2] & 70           & 1.0$\times$10$^6$     & 1.06--1.74 \\
\hline
\end{tabular}
\caption{Narrowband surveys for \lya\ at $z = 5.7$}
\label{tab:otherSurveys}

\footnotetext[1]{CADIS is based on imaging with a tunable Fabry-Perot
  interferometer scanning at equally spaced wavelength steps
  \citep{Hippelein03}.}

\footnotetext[2]{An additional encompassing mediumband filter was used
  here.}

\end{center}
\end{minipage}
\end{table*}

\begin{figure}[tbp]
\centering
\includegraphics[width=\columnwidth, trim=13pt 5pt 18pt 18pt, clip]{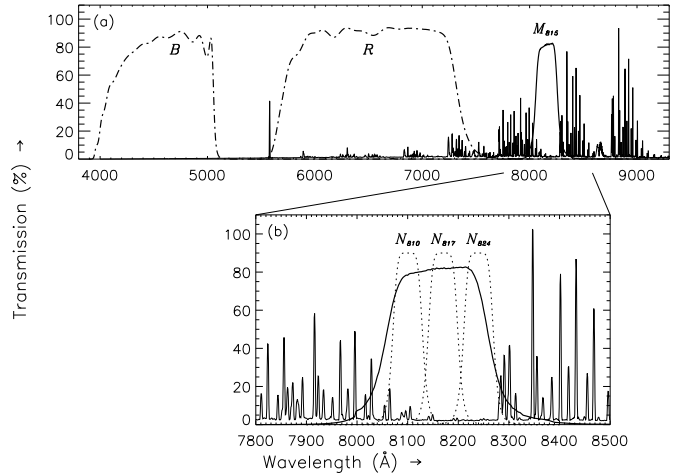}
\caption{Filter set used for the WFILAS Survey. {\bf a)} The broadband
  $BR$ and intermediate \med\ (815/22\,nm) passbands. {\bf b)} The
  \med\ intermediate passband ({\em solid} line) and three
  purpose-built narrowbands \na\ (810/7), \nb\ (817/7) and \nc\
  (824/7), shown over the wavelength region targeted for \lya\ line
  detections. The transmission curves of the narrowband filters are
  for illustrative purposes only. The OH night sky line background is
  also shown.}
\label{fig:filters}
\end{figure}

The sky area surveyed by the WFILAS is $\sim$0.74\,sq.\ degree. We
observed three fields in broadbands $B$, $R$ and in an intermediate
width filter centred at 815\,nm encompassing three narrowband filters
(Fig.~\ref{fig:filters}). The adoption of an additional intermediate
width filter encompassing the multiple narrowband width filters is a
novel approach compared to previous narrowband surveys. The
application of the intermediate band filter enables us to drastically
reduce the number of spurious detections in the narrowband filters.
The narrow width of the narrowband filters ({\it FWHM}\,=\,7\,nm) gives a
prominent appearance to emission line objects. Furthermore, the three
chosen fields are spread across the sky to enable us to average out
variations in cosmic variance. Our search has covered one of the
largest co-moving volumes compared to other surveys.
Table~\ref{tab:otherSurveys} compares WFILAS with other published
surveys.

The observations were taken with the Wide Field Imager
\citep[WFI;][]{Baade99} on the ESO/MPI 2.2\,m telescope at the Cerro
La Silla Observatory, Chile. The data were taken over 65 separate
nights from 2001 January 19 to 2003 December 1. The WFI is a mosaic of
eight ($4 \times 2$) 2k $\times$ 4k CCDs arranged to give a field of
view of 34\am\ $\times$ 33\am. The pixels are 0\farcs238 on a side.

As WFILAS was planned as joint project of ESO Santiago and the
COMBO-17 team at MPIA Heidelberg, three fields were selected to
overlap with the COMBO-17 survey, i.e. their extended Chandra Deep
Field South (CDFS), SGP (South Galactic Pole) and S11 fields. The
coordinates of the field centres and the exposure times in each of the
filters for each field are given in Table~\ref{tab:filters}. All three
fields are at high Galactic latitude ($|b| > 54^\circ$) and have
extinctions less than $E(B-V)$\,=\,0.022 mag \citep{Schlegel98}.

\begin{table*}[!htbp]
\begin{minipage}[htb]{\textwidth}
\begin{center}
\renewcommand{\thefootnote}{{\it\alph{footnote}}}
\begin{tabular}{ccccccccccc}
\hline\hline
Filter                          & Passband/ & \multicolumn{3}{c}{{\bf CDFS field}}    & \multicolumn{3}{c}{{\bf S11 field}}     & \multicolumn{3}{c}{{\bf SGP field}}    \\
                                & {\it FWHM}      & \multicolumn{3}{c}{03\h~32\m~25\fs134}  & \multicolumn{3}{c}{11\h~42\m~59\fs933}  & \multicolumn{3}{c}{00\h~45\m~55\fs024} \\
                                & (nm)      & \multicolumn{3}{c}{$-27$\d~48\am~49\farcs75} & \multicolumn{3}{c}{$-01$\d~42\am~46\farcs44} & \multicolumn{3}{c}{$-29$\d~34\am~55\farcs05}\\
\hline                                                                                                                                                                           
                                &           & ($a$)             & ($b$)     & ($c$)   & ($a$)             & ($b$)  & ($c$)      & ($a$)             & ($b$)     & ($c$)  \\
Narrowband \na\                 & 810/7     & 48.0              & 0.57        & 0.79    & 44.4              & 0.55     & 0.80       & 31.5              & 0.87        & 1.03   \\
Narrowband \nb\                 & 817/7     & 41.1              & 0.55        & 0.79    & 79.9              & 0.53     & 0.92       &  0.0              & -           & -      \\
Narrowband \nc\                 & 824/7     & 41.0              & 0.72        & 0.80    & 43.5              & 0.81     & 0.87       & 42.8              & 0.62        & 0.89   \\
Mediumband \med\footnotemark[1] & 815/20    & 52.7              & 0.29        & 0.85    & 33.3              & 0.38     & 0.88       & 18.9              & 0.41        & 0.90   \\
Broadband $B$\footnotemark[1]   & 458/97    &  5.0              & 0.07        & 1.09    &  9.4              & 0.07     & 0.98       & 10.0              & 0.14        & 1.22   \\
Broadband $R$\footnotemark[1]   & 648/160   & 15.1              & 0.05        & 0.75    & 21.2              & 0.07     & 0.75       & 21.5              & 0.07        & 0.76   \\
\hline\\                                                                                                                                                                           
\end{tabular}
\caption{WFILAS fields, filter set exposure times and detection
  limits. The entries under each field heading list: (a) the
  total exposure time (ks), (b) the flux for a 2$\sigma$
  detection on 6 pixel diameter aperture (\uJy) and (c) the final
  seeing (\as), in each filter.}
\label{tab:filters}
\footnotetext[1]{Broadband $B$ and $R$ and part of the intermediate
  band \med\ taken from the COMBO-17 survey \citep{Wolf04}}
\end{center}
\end{minipage}
\end{table*} 

We employ standard broadband $B$ and $R$ filters. The intermediate
band ({\it FWHM} = 22\,nm) observatory filter is centred at 815\,nm. The
three custom made narrowband ({\it FWHM} = 7\,nm) filters are centred at
810\,nm, 817\,nm and 824\,nm. The transmission profiles of the filters
are shown in Fig.~\ref{fig:filters}. The intermediate and narrowband
filters are designed to fit in the atmospheric 815\,nm OH-airglow
window, where the brightness of the sky background is low and hence
favourable to detect \lya\ emission at redshift $\sim$5.7. The data
taken with the intermediate band filter confirm detections of the
\lya\ line in one of the narrowband filters. The broadband $B$ and $R$
data, which were taken from the COMBO-17 survey \citep{Wolf04}, are
used to confirm the absence of continuum blueward of the \lya\ line
and to avoid sample contamination by lower redshift emission line
galaxies (e.g. \ha\ at $z \sim 0.24$, or \oii\ at $z \sim 1.2$).

To establish the photometric zero-point of the intermediate and
narrowband filters two spectrophotometric standard stars
\citep[LTT3218 and LTT7987;][]{Bessell99} were observed.

Between \mbox{10--50} exposures were taken for each intermediate and
narrowband filter for each field. The exposure times varied between
1000 and 1800\,sec per frame, with a typical exposure time of around
1600\,sec. All frames are background-limited despite the low night sky
emission in this spectral region. The median, first and last decile of
both seeing and background are given in Table~\ref{tab:imgquality}.

\begin{table}[!htbp]
\centering
\begin{tabular}{lcrrrrrr}
  \hline\hline\\
  Filter  & No. of & \multicolumn{3}{c}{Background ($\mu$Jy/$\Box$\arcsec)} & \multicolumn{3}{c}{Seeing (\arcsec)}\\
  & Frames & 10\%  &  50\% &  90\%          &  10\% & 50\% & 90\%\\
  \hline\hline\\
  \na\    &  92    & 17    & 27    & 36             & 0.65  & 0.79 & 1.12\\
  \nb\    &  75    & 19    & 30    & 41             & 0.64  & 0.84 & 1.16\\
  \nc\    &  77    & 17    & 27    & 36             & 0.63  & 0.80 & 1.10\\
  \med\   &  80    & 17    & 22    & 33             & 0.65  & 0.83 & 1.09\\
  \hline\hline\\
\end{tabular}
\caption{The median, first and last decile of background and seeing for the
  WFILAS narrowband imaging for all three fields combined. One
  pixel corresponds to 0\farcs238.}

\label{tab:imgquality}
\end{table}

\section{Data Reduction}
\label{sec:datared}
The data were processed with standard IRAF\footnote{IRAF is
  distributed by the National Optical Astronomy Observatories, which
  are operated by the Association of Universities for Research in
  Astronomy, Inc., under cooperative agreement with the National
  Science Foundation.} routines (MSCRED TASK) and our own specially
designed scripts. The initial steps in the reduction process consist
of removing the zero level offset with bias frames, normalising
pixel-to-pixel sensitivity differences with twilight flatfield frames
and removal of fringes with fringe frames. During these steps, the 8
CCDs that make up a single WFI image are treated independently. These
processes are described in detail below.

Normally, the overscan region of the science frames can be used to
remove the zero level offset. However, it was noticed that the bias
frames contained significant intermediate scale structure (10-30
pixels). To remove this, bias frames were taken on every day of our
observations and averaged into a bias frame for that day. In order to
minimise the noise added to the data by subtracting the bias, the bias
frames were smoothed by 5 pixels and 30 pixels in horizontal and
vertical direction of the CCDs, respectively, and subsequently
medianed. The structures are stable over periods of several months.
Therefore, it was possible to use bias frames from different nights
without degrading the quality of the data.

Typically, five twilight flatfield frames were taken in one night for
one or more filters. The frames were medianed and the science data was
divided by the median. Hence pixel-to-pixel sensitivity differences
were removed. The structure in the individual flatfield frames was
stable over a period of several weeks. Frames taken on different
nights could thus be reused. Any differences between flatfield frames
were due to the appearance or disappearance of dust features, or large
scale illumination differences. The differences rarely amounted to
more than a few percent.

The raw data in the intermediate and narrowband filters show fringe
patterns with amplitudes of up to 10\% which was only partially
removed after the data had been flatfielded. To entirely remove the
fringe pattern, we subtracted a fringe frame created from
\mbox{10--30} science frames. The fringing is very stable over time,
so we were able to use data spanning several months. Certain science
frames still show fringe patterns because they are contaminated by
either moonlight or twilight. Residual differences in the level of the
background between the different CCDs were removed by subtracting the
median background level from each CCD.

To produce the final deep images we only used images with a seeing of
less than 5 pixels (=\,1\farcs2) and without significant residual
fringing. To make the combining of the images possible, we had to
apply an astrometric correction based on stars from the USNO CCD
Astrograph Catalogue 2 \citep[UCAC2;][]{Zacharias04} in the three
observed fields. The frames have a set pixel scale of 0\farcs238
pixel$^{-1}$ with North up and East left. The images were weighted
according to their exposure time and combined using the IRAF
``mscstack'' routine rejecting deviant pixels. Table~\ref{tab:filters}
summarises the depth, image quality and total exposure time, for each
coadded frame.

\section{Sample Selection and Completeness}
\label{sec:selection}
\subsection{Photometry and Noise Characteristics}
Initial source catalogues were created for each of the 8 narrowband
images. Each catalogue contains the photometry for the sources in all
6 filters. We used the \sext\ source detection software \citep[version
2.3.2, double image mode;][]{Bertin96}. Sources were selected when at
least 5 pixels were 0.8$\sigma$ above the noise level in the
narrowband image used for detection. The photometry was measured in
two apertures, 6 and 10 pixels in diameter (=\,1\farcs4 and 2\farcs4,
respectively). The 6 pixel aperture was used to maximise the
signal-to-noise of the flux of the objects, while the larger 10 pixel
aperture was used for the more accurate determination of the total
flux and hence the star formation rate.

Some authors have found that \sext\ underestimates flux uncertainties
\citep{Feldmeier02,Labbe03}. \sext\ estimates the uncertainties using
various assumptions that are often not valid (e.g. perfect
flatfielding, perfect sky subtraction). The pixel-to-pixel noise in
our data is slightly correlated because the scatter in the counts
summed in 6 pixel apertures is about 10\% higher than what one would
derive from the measured pixel-to-pixel RMS.

\begin{figure}[tbp]
  \centering
  \includegraphics[width=\columnwidth, trim=16pt 10pt 19pt 19pt, clip]{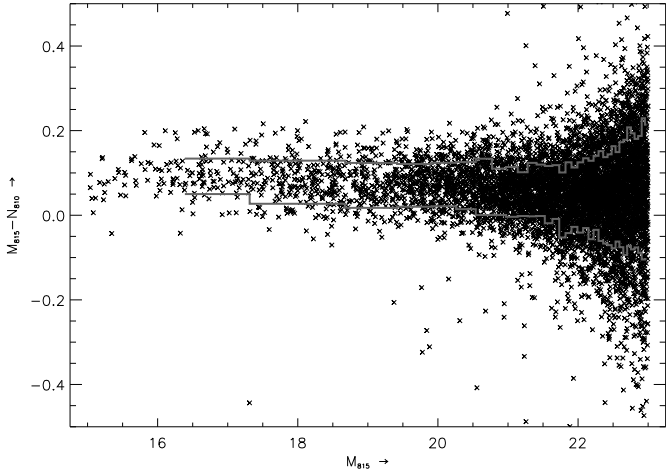}
  \caption{\mminn\ colour as seen in the S11 field with the \na\
    filter as the detection image. The mean colour term is
    $\sim$0.05. The heavy bold lines indicate the 68.3\% interval of
    objects colours closest to the mean colour in each bin. Each bin
    contains 200 data points. The new uncertainty is inferred from
    this interval.}
  \label{fig:trumpet}
\end{figure}

We devised a method to correct the uncertainties given by \sext\ to
their true values as follows. First, sources with flux in all filters
and their \med\ magnitude between 16 and 23 were selected. Sources
brighter than \med\,=\,16 are typically saturated, while those fainter
than \med\,=\,23 are incomplete (see Sect.~\ref{sec:completeness}
for a further discussion of incompleteness). The \mminn\ colour (where
$N$ is any of narrowband filters \na, \nb, or \nc) is the same for any
flat continuum source. Therefore, the spread in the \mminn\ colour
will be the same as the true flux uncertainty from the two
contributing filters. Next, the sources were binned into 200-source
bins based on their \med\ magnitude. In Fig.~\ref{fig:trumpet} we plot
the \mminn\ colour versus the \med\ magnitude of one of our S11
catalogues. Mean values for the \mminn\ colour, \med, $N$ magnitude
and the mean of the \sext\ uncertainty were calculated for each bin.
The uncertainty in the colour for each object was determined by adding
the uncertainty of \med\ and $N$ in quadrature ($\sigma^2_{col} =
\sigma^2_M + \sigma^2_N$). The interval in which 68.3\% of the
objects were closest to this mean colour was used to infer the actual
1$\sigma$ colour uncertainty. We assumed that the ratio between the
old uncertainties $\sigma_M$ and $\sigma_N$ was the same for the new
uncertainties $\sigma_M'$ and $\sigma_N'$. We related between the new
and old uncertainty in the intermediate and narrowband flux using the
function $\sigma_{\rm filter}' = \sqrt{a^2 + (b \sigma_{\rm
    filter})^2}$, where $a$ is the zero-offset for the uncertainty in
the flux of bright sources and $b$ is the ratio between the new and
old uncertainty for the flux of the faintest sources. The parameters
$a$ and $b$ correspond to imperfections in the photometry and wrongly
assumed background by \sext, respectively.

Typically, the correction factors are moderate (between
$\sim$30$-$50\%) for the faint sources in the catalogues. Even
though the correction factors are moderate, we assume that the
corrections for the uncertainties in the broadband $B$ and $R$ are
irrelevant, since they are used in a different way than the
intermediate and narrowband images (see Sect.~\ref{sec:criteria}).

\subsection{Selection criteria}
\label{sec:criteria}

The following four criteria were applied to select our candidate LAEs
from the eight initial source catalogues:
\begin{enumerate}
\item the narrowband image used as the detection image must have the
  most flux of all the narrowband images and the source must have a
  4$\sigma$ detection or better;
\item the narrowband image with the least flux needs to be a
  non-detection, i.e. less than 2$\sigma$;
\item there must be at least a 2$\sigma$ detection in the
  intermediate band image;
\item none of the broadband images, i.e. neither $B$ nor $R$, must
  have a detection above 2$\sigma$.
\end{enumerate}
Table~\ref{tab:filters} contains the values of the 2$\sigma$
detection thresholds of the images used for the 6 pixel aperture. In
total 33 candidates were selected using the above criteria. Visual
inspection showed that 26 sources arose from artefacts of which the
vast majority were out-of-focus ghost rings from bright stars. The
final sample contains seven candidate LAEs.

We note here the importance of the usage of the intermediate band
filter. If we were to reapply all the criteria except for criterion 3,
i.e. we do not use the intermediate band images, we would obtain 284
candidates instead of the 33 for visual inspection.

\begin{figure*}[!htbp]
  \begin{center}
  \includegraphics[width=\textwidth]{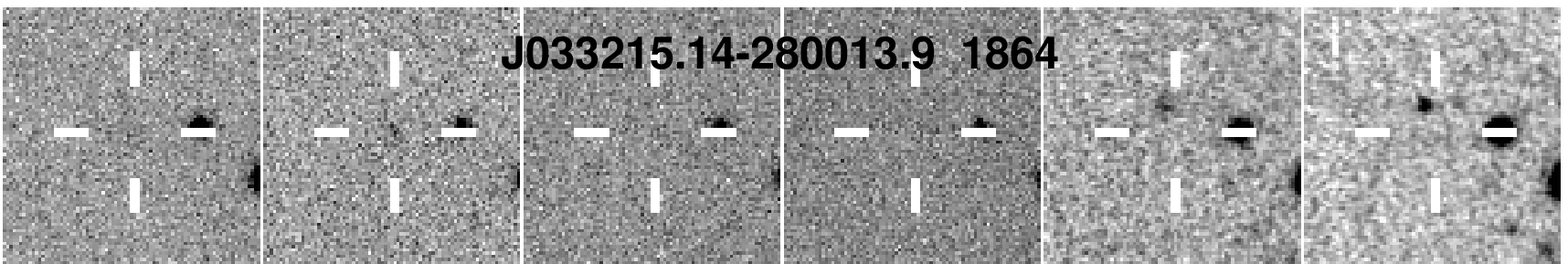}
  \includegraphics[width=\textwidth]{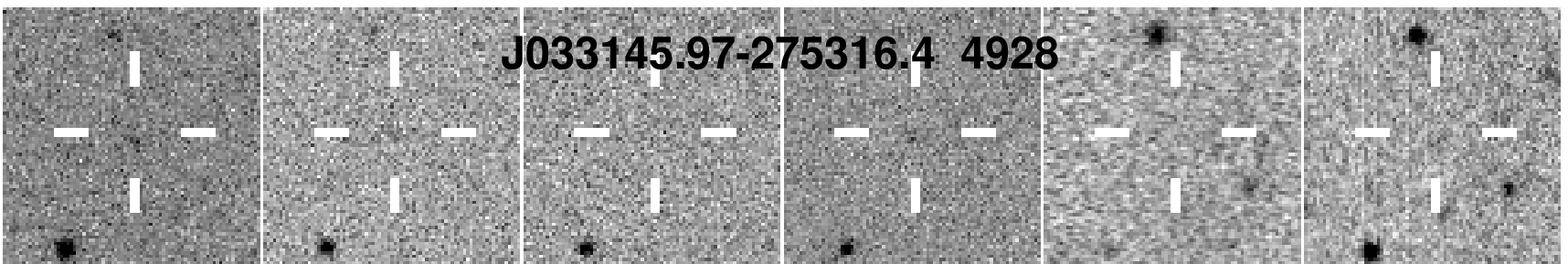}
  \includegraphics[width=\textwidth]{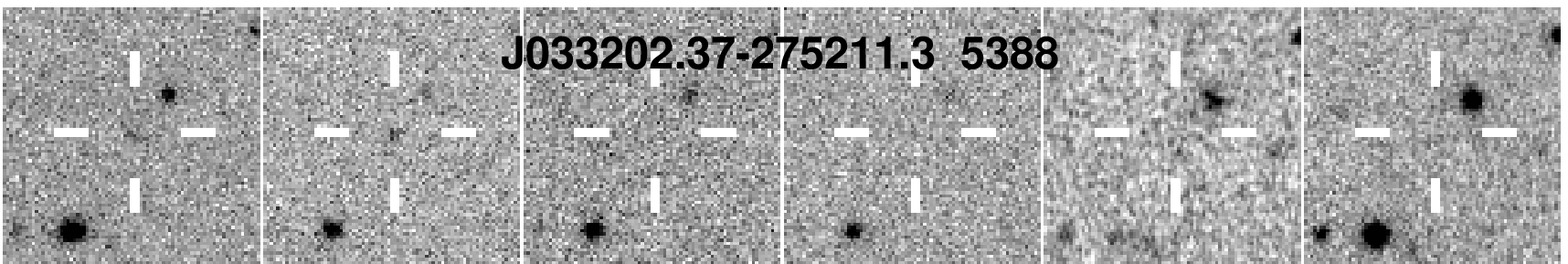}
  \includegraphics[width=\textwidth]{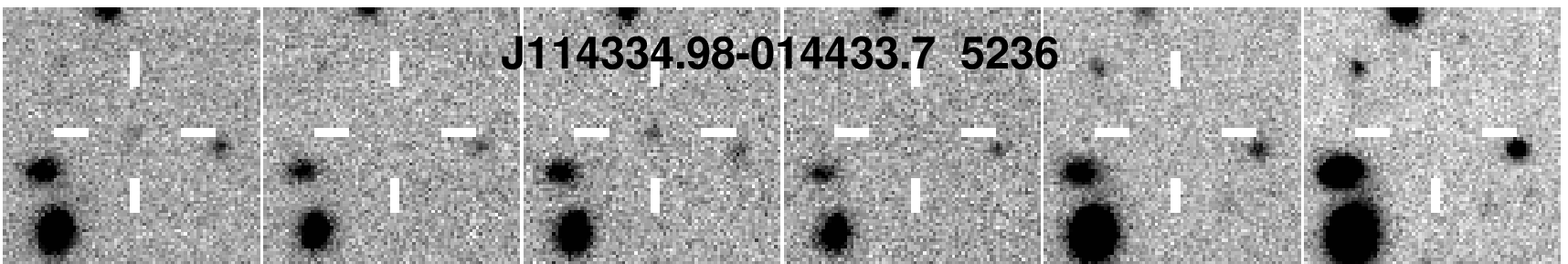}
  \includegraphics[width=\textwidth]{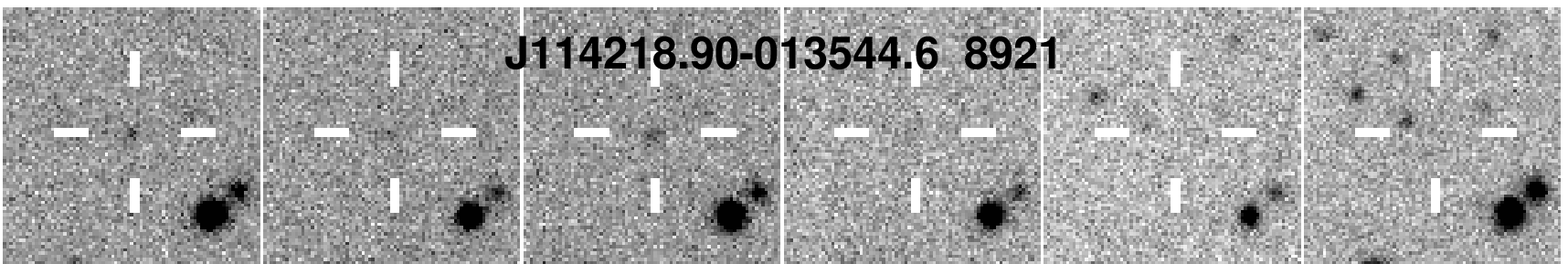}
  \includegraphics[width=\textwidth]{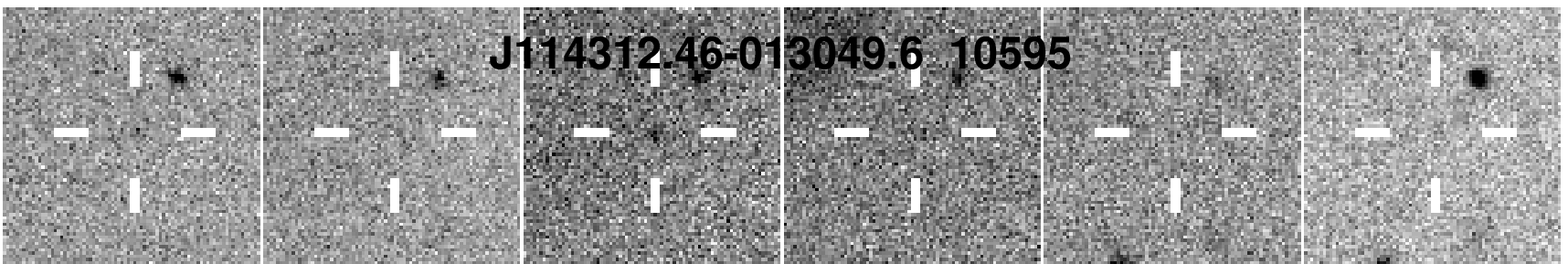}
  \includegraphics[width=\textwidth]{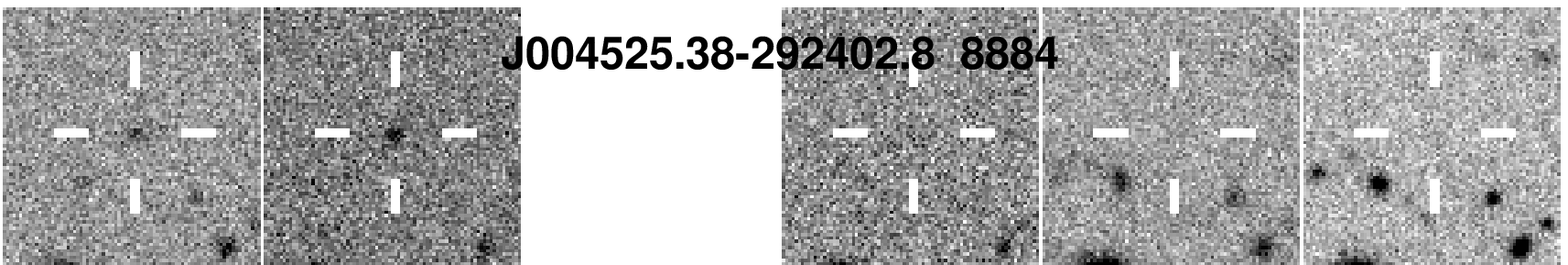}
\end{center}
  \makebox[0pt][l]{\hspace{0.045\textwidth} \huge \med}
  \makebox[0pt][l]{\hspace{0.210\textwidth} \huge \na}
  \makebox[0pt][l]{\hspace{0.370\textwidth} \huge \nb}
  \makebox[0pt][l]{\hspace{0.535\textwidth} \huge \nc}
  \makebox[0pt][l]{\hspace{0.705\textwidth} \huge $B$}
  \makebox[0pt][l]{\hspace{0.870\textwidth} \huge $R$}
  \caption{Thumbnails of each region in which the candidate LAEs
    appears. The thumbnails cover a 19\as$\times$19\as\ region with a
    pixel scale of 0\farcs238\,pixel$^{-1}$ and North is up and East
    to the left. From left to right are the filters \med, \na, \nb,
    \nc, $B$ and $R$.}
\label{fig:thumbs}
\end{figure*}

The {\it AB}-magnitudes, derived line fluxes and luminosities for the
candidates are shown in Table~\ref{tab:cands}. To convert between
{\it AB}-magnitudes and line flux in \lineunits\ we use the following
relation:
\begin{equation}
  F_{\rm line} = 3\times\pow{18}\,\pow{-0.4(m_{AB} + 48.590)}\,\frac{\Delta \lambda}{\lambda_c^2}
\end{equation}
where $\Delta \lambda$ and $\lambda_c$ are the {\it FWHM} and the central
wavelength of the narrowband filter in \AA, respectively, and $m_{AB}$
the {\it AB}-magnitude of the object. In Fig.~\ref{fig:thumbs} the
thumbnails of the seven candidate LAEs at $z\sim5.7$ are shown. We
defer a more detailed discussion about the sample properties to
Sect.~\ref{sec:candcat}.

\subsection{Completeness corrections}
\label{sec:completeness}
\begin{figure}[tbp]
  \includegraphics[width=\columnwidth, trim=34pt 8pt 18pt 18pt, clip]{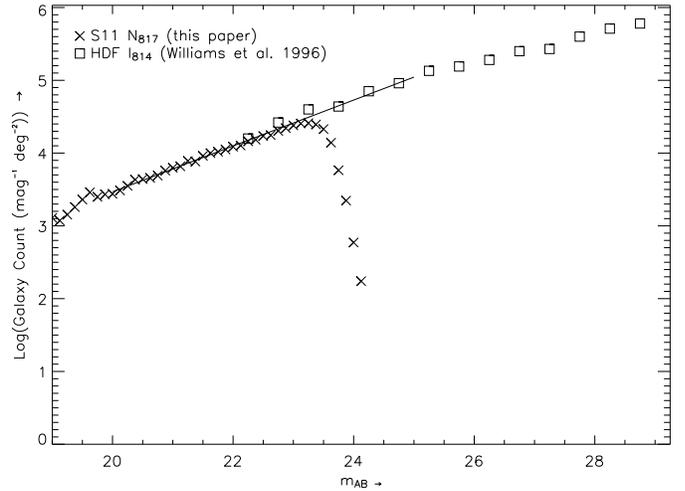}
  \caption{Galaxy counts as a function of {\it AB}-magnitude for the \nb\
    filter in the S11 field. Plotted are the $N_{817}$ source counts
    of the S11 field ({\it crosses}) together with the $I_{814}$
    galaxy counts of the Hubble Deep Field ({\it squares}). The solid
    line indicates the fitted linear relation between the magnitude
    and galaxy count.}
  \label{fig:galcount}
\end{figure}

From the Hubble Deep Field (HDF) galaxy number-count data for the {\it
  F}814{\it W} filter \citep{Williams96} we computed completeness
corrections for our eight source catalogues. The HDF counts are
determined over the magnitude range $I_{814}$\,=\,22\,$-$\,29, and
agree well with our galaxy counts over all narrowband filters in the
range $N$\,=\,22\,$-$\,24. Figure~\ref{fig:galcount} shows the counts
for the {\it F}814{\it W} filter in the HDF and for the \nb\ filter in
the S11 field.  Figure~\ref{fig:galcount} also shows the linear fit
used as the basis for the calculation of the detection completeness.
The fit is done to the combined number count data over two intervals:
\nb\,=\,[20,\,22.5], where the WFILAS counts are complete, and
$I_{814}$\,=\,[22.5,\,25], where the HDF counts are linear.

\begin{figure}[tbp]
  \includegraphics[width=\columnwidth, trim=20pt 10pt 19pt 19pt, clip]{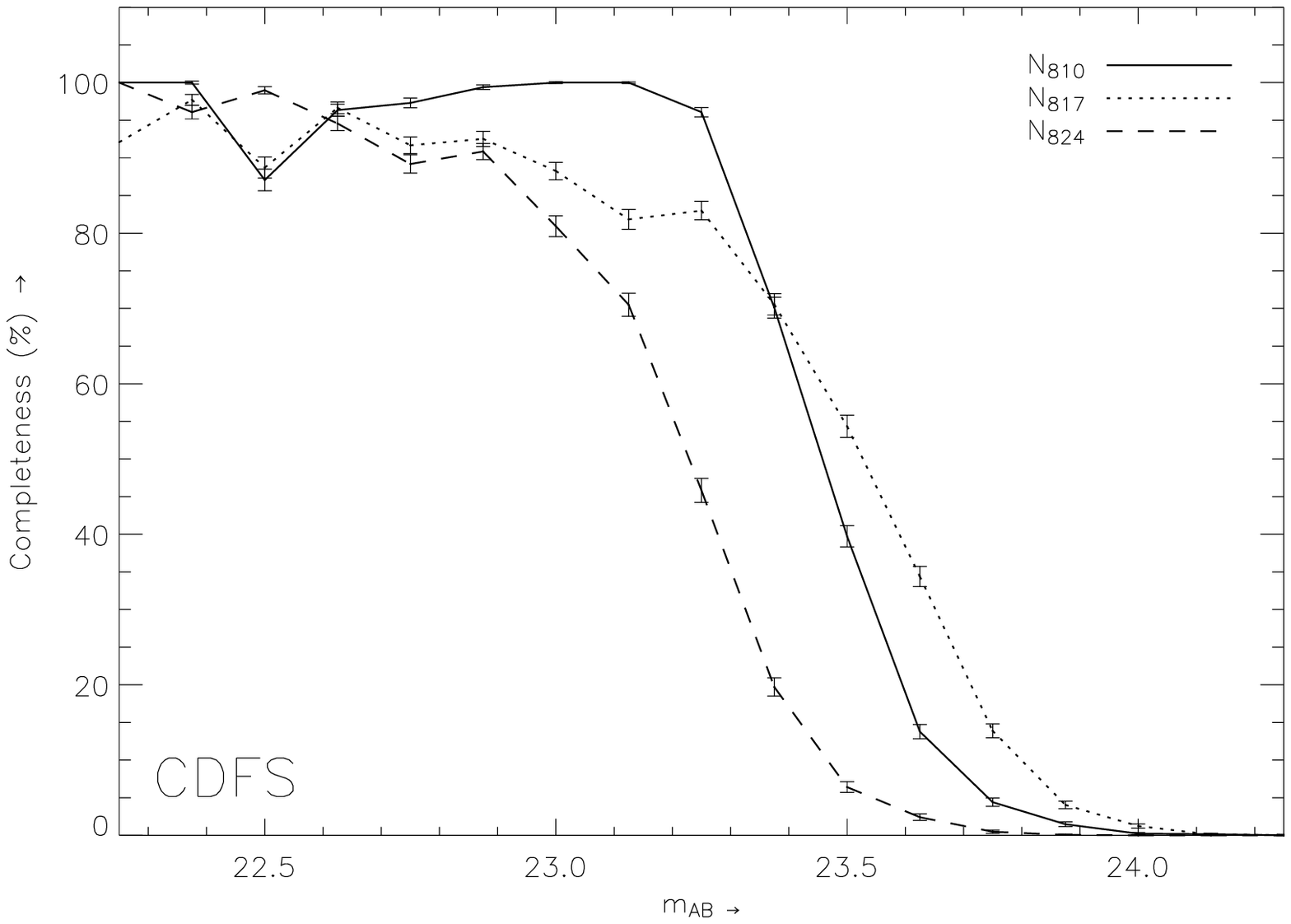}
  \includegraphics[width=\columnwidth, trim=20pt 10pt 19pt 19pt, clip]{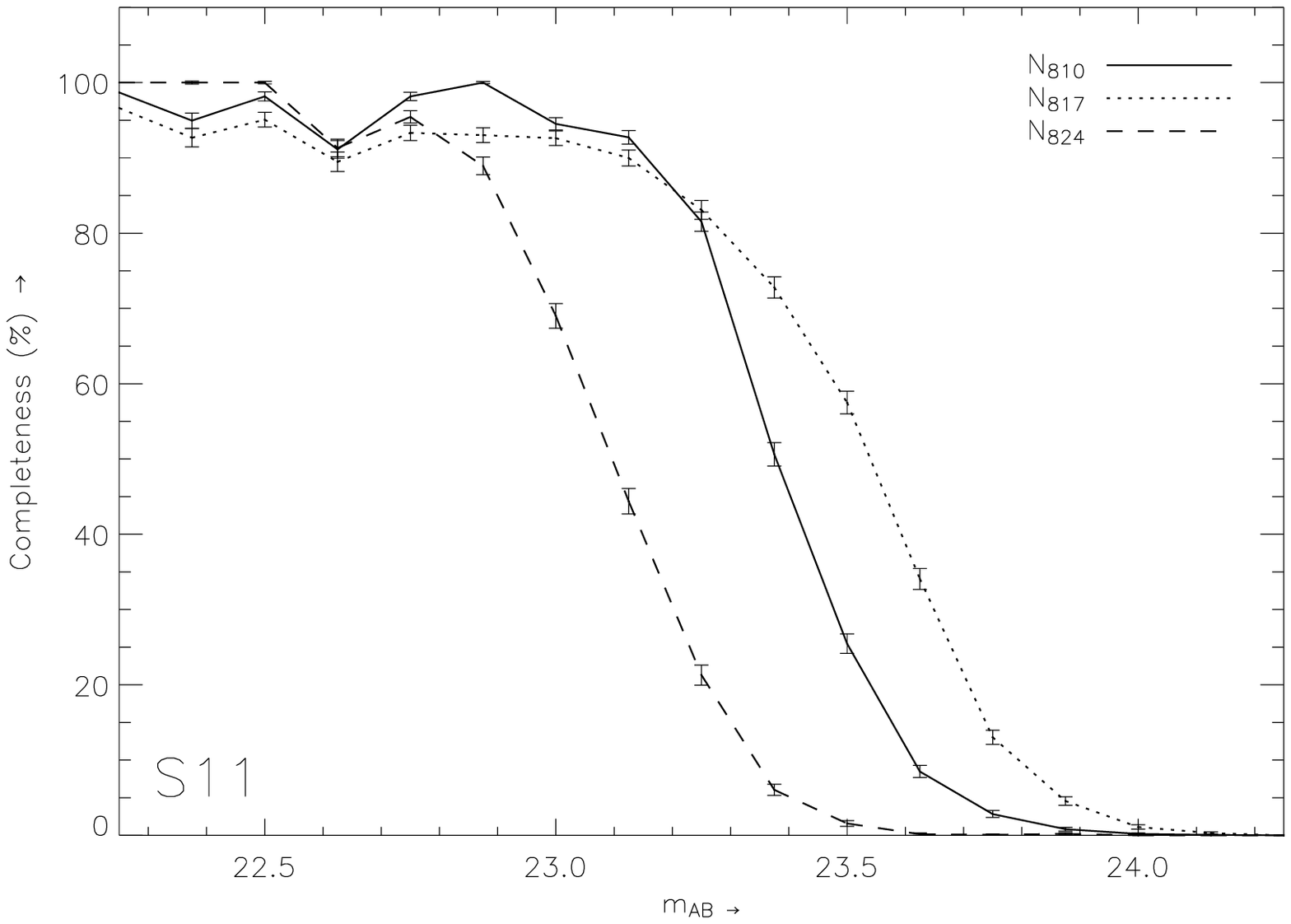}
  \includegraphics[width=\columnwidth, trim=20pt 10pt 19pt 19pt, clip]{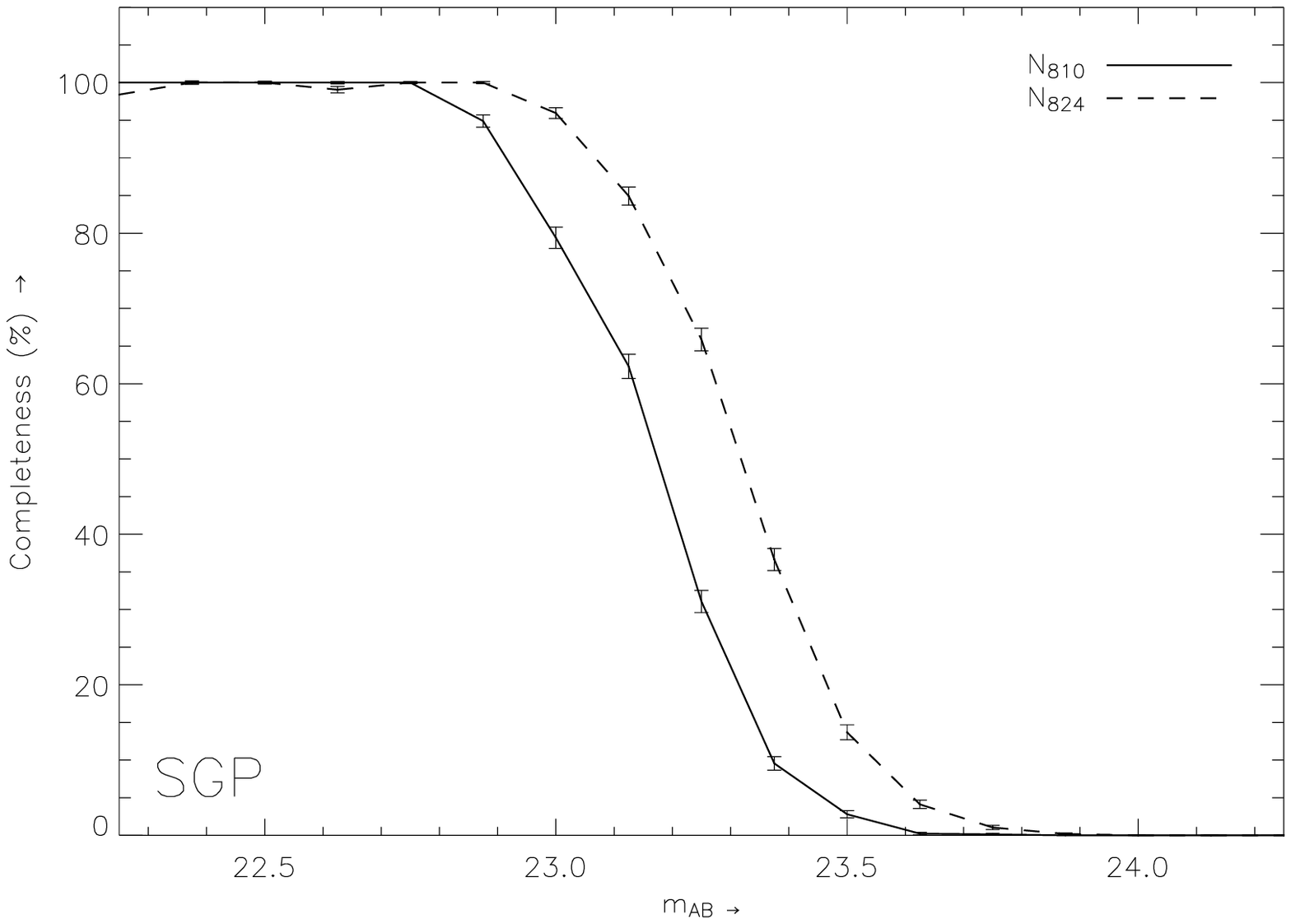}
  \caption{Detection of completeness as function of magnitude derived
    from the galaxy density-magnitude relation as described in
    Sect.~\ref{sec:selection}. From top to bottom the CDFS, S11 and
    SGP fields are shown. No \nb\ data are available for the SGP
    field.}
  \label{fig:completeness}
\end{figure}

Detection completeness is defined as the ratio of WFILAS sources to
the number expected from the number-count relation.
Figure~\ref{fig:completeness} shows the derived detection completeness
for each filter-field combination used for WFILAS. The differences are
mainly due to unequal exposure times, although filter throughput and
image quality also play a role. These could explain the overall lower
sensitivity of the \nc\ filter, as can be inferred from
Fig.~\ref{fig:completeness}. Additionally, we correct for detection
completeness arising due to the intermediate band selection
criterion. We constructed a noise image by stacking the intermediate
band images without registering. The completeness is defined as the
rate of recovery of artificially inserted objects.

\begin{figure}[tbp]
  \includegraphics[width=\columnwidth, trim=33pt 7pt 18pt 18pt, clip]{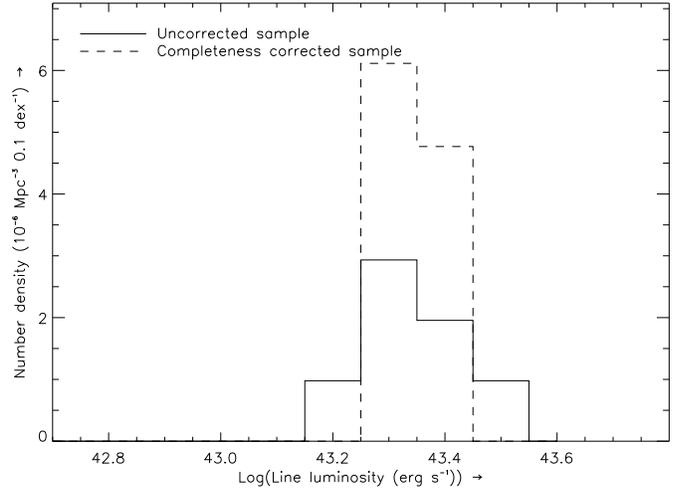}
  \caption{Line luminosity distribution of the candidate LAEs. Two
    samples are indicated: all the candidates, but not corrected for
    completeness ({\it solid}) and the candidates in the complete
    sample, i.e. candidates of the four deepest narrowband images with
    a magnitude cut-off at 50\% completeness of the worst of these
    four images ({\it dashed}).}
  \label{fig:lumdist}
\end{figure}

Given the different sensitivities of each filter-field combination, we
define a homogeneous subsample of our initial candidate sample, using
the candidates from our four most sensitive field-filter combinations.
We call this our ``complete'' sample (4 of the 7 LAEs; marked in
Table~\ref{tab:cands}), because once defined, we use the curves in
Fig.~\ref{fig:completeness} to correct the detected candidate numbers
for incompleteness, in contrast to our initial ``incomplete'' sample
(all 7 LAEs). The purpose of the subsample is that it lies within a
uniform flux limit. Figure~\ref{fig:completeness} shows that our four
best filter-field combinations consist of the \na\ and \nb\ filters in
both the CDFS and S11 fields. These four field-filter combinations
reach at least 50\% completeness at $M_{\rm AB}$\,=\,23.38, or
5.1$\times$10$^{-17}$\,\lineunits. We take this as the flux limit of
our complete sample. As such, the number density derived from the
complete sample is a more accurate measure of the density of sources
down to the nominated flux limit than the number density of the
incomplete sample. Figure~\ref{fig:lumdist} shows the luminosity
distribution of the complete sample alongside our initial candidate
list, which we call the ``incomplete'' sample. It shows that in using
completeness corrections our detected source density is up by 50\%.

\section{$z\sim5.7$ Candidate LAE Catalogue}
\label{sec:candcat}

In the previous Sect. we introduced two sets of candidate LAEs: the
full (but incomplete) sample of seven candidate LAEs and a subsample
thereof, complete to $F_{lim}$\,=\,5.1$\times$10$^{-17}$\,\lineunits\
(the complete sample). The flux limit of the incomplete sample is
almost twice the limit of the complete sample
(3.4$\times$10$^{-17}$\,\lineunits).

\begin{figure}[tbp]
  \centering
  \includegraphics[width=\columnwidth, trim=15pt 9pt 19pt 20pt, clip]{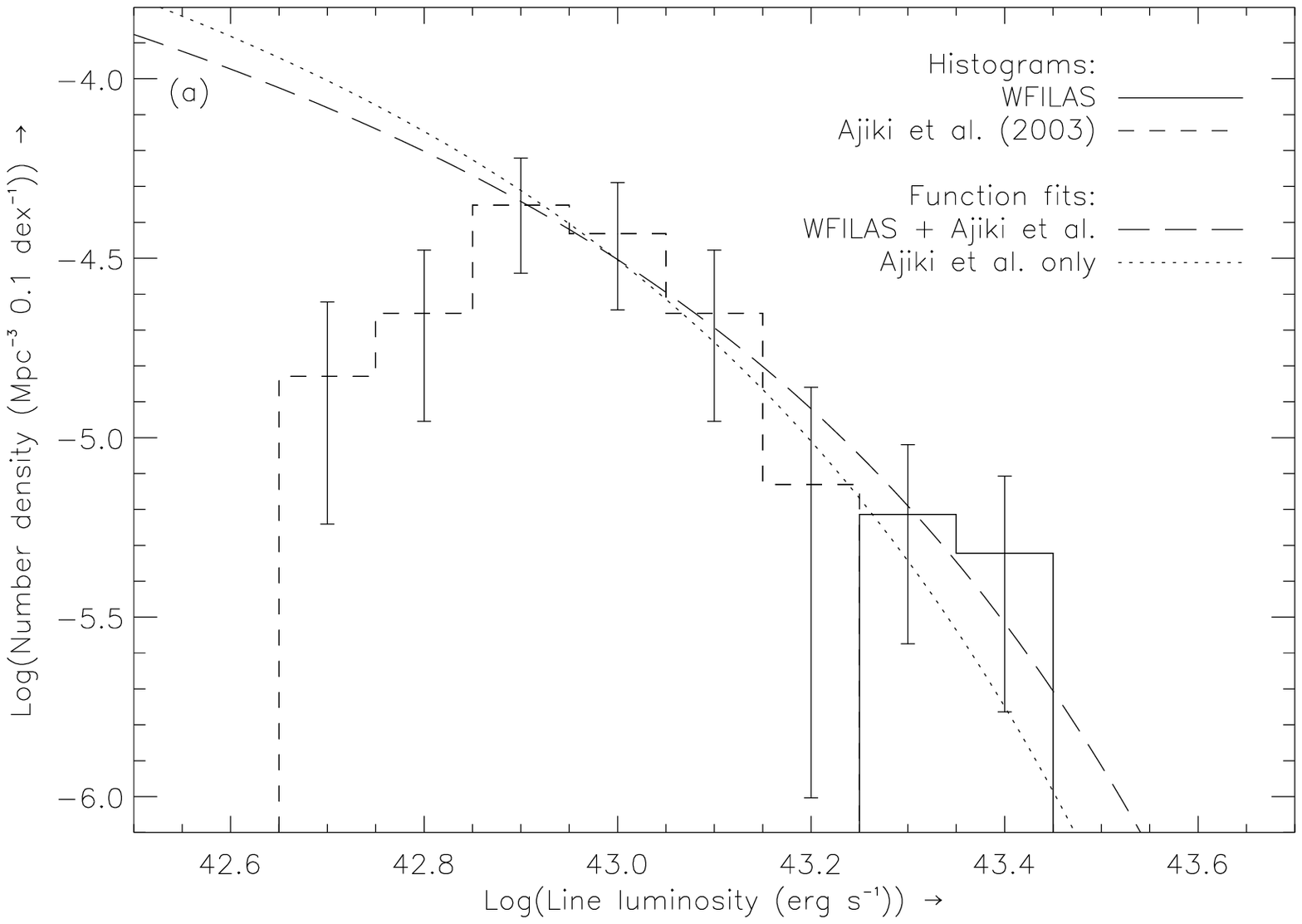}
  \includegraphics[width=0.75\columnwidth, trim=15pt 9pt 19pt 20pt, clip]{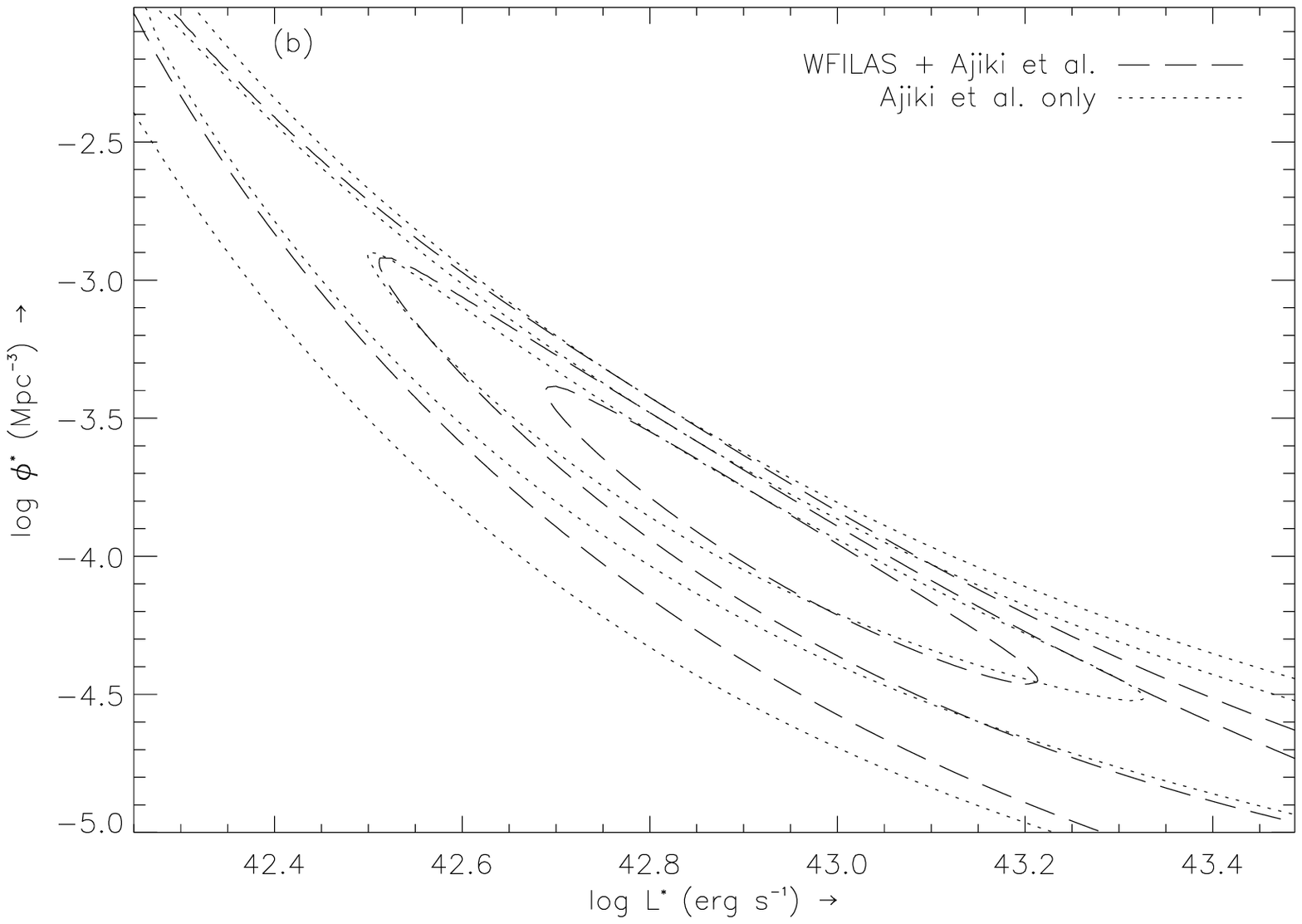}
  \caption{{\bf a)} Line luminosity distribution of the complete
    sample of candidate LAEs ({\it solid histograms}) together with
    the candidates from \citet{Ajiki03} ({\it dashed histograms}).
    Both samples are corrected for completeness. The errorbars are
    derived using Poisson statistics. Furthermore, two Schechter
    function fits are indicated: one to the combined WFILAS and Ajiki
    sample ({\it dotted}) and one to Ajiki sample only ({\it long
      dashed}). {\bf b)} The 68.3\%, 95.4\% and 99.7\% confidence
    limits for the fitting parameters $L^*$ and $\phi^*$. See text for
    details.}
  \label{fig:complumdist}
\end{figure}

\begin{table}[tbp]
  \centering
  \begin{tabular}{cc}
    \hline
    \hline
    Log(L (\ergs)) & Log($\Phi$ (Mpc$^{-3}$ 0.1\,dex$^{-1}$))\\
    \hline
    42.7 & -4.83$^{+0.21}_{-0.41}$\\
    42.8 & -4.65$^{+0.18}_{-0.30}$\\
    42.9 & -4.35$^{+0.13}_{-0.19}$\\
    43.0 & -4.43$^{+0.14}_{-0.21}$\\
    43.1 & -4.65$^{+0.18}_{-0.30}$\\
    43.2 & -5.13$^{+0.27}_{-0.87}$\\
    43.3 & -5.21$^{+0.19}_{-0.36}$\\
    43.4 & -5.32$^{+0.21}_{-0.44}$\\
\hline
  \end{tabular}
  \caption{Number density of LAEs per luminosity bin as
    indicated in Fig.~\ref{fig:complumdist}a.}
\end{table}

To examine the luminosity distribution of our sample we use the
Schechter function \citep{Schechter76}, as it is a good representation
of the data at bright luminosities. From this, the luminosity density
$\mathcal{L}$ of a distribution with a limiting luminosity $L_{lim}$
is given by
\begin{equation}
\mathcal{L}(L \ge L_{lim}) = \phi^* L^* \Gamma(\alpha+2, L_{lim}/L^* ),
\label{eq:lumdens}
\end{equation}
where $\alpha$ and $\phi^*$ represent the slope of the faint end of
the Schechter function and the normalisation constant of the galaxy
density, respectively. $\Gamma$ is the incomplete gamma-function.
Currently, the luminosity function for LAEs at $z\sim5.7$ is poorly
defined and authors commonly adopt either one or two of the three
parameters from low redshift surveys to calculate the third.

We examine the influence of non-detections of bright ($L \gtrsim L^*$)
LAEs for the total \lya\ luminosity density by employing the same
method as \cite{Ajiki03}, another narrowband imaging survey aimed at
finding LAEs at $z\,\sim\,5.7$. In the interest of comparison, we
follow \citeauthor{Ajiki03} exactly and adopt the \citet{Fujita03}
values for $\alpha$ (\mbox{-1.53}) and $\phi^*$
(10$^{-2.62}$\,\perMpc3). Their approach was to solve
Eq.~(\ref{eq:lumdens}) for $L^*$, instead of fitting a Schechter
function. Fixing $\phi^*$ and allowing $L^*$ and $\alpha$ to vary
imposes a strong prior on the final fit, it allows us to compare
directly to the results of \citeauthor{Ajiki03} by preserving their
method. The luminosity density $\mathcal{L}$ was calculated by summing
the luminosity of all candidates (corrected for completeness) and
divided by the corresponding survey volume. With the given survey
limits the equation can be solved for $L^*$. Equation~(\ref{eq:lumdens})
yields the total luminosity density when $L_{lim}=0$. We have done
this for three cases: for the candidates of \citeauthor{Ajiki03} (case
A), the complete sample of our candidates (case B) and a combined
sample of these two surveys (case C). For our complete sample we
derive a higher $L^*$ (+0.12\,dex; case B) than \citet[][case
A]{Ajiki03} which implies an increase of the luminosity density
$\mathcal{L}$ of $\sim$30\%. If we scale the luminosity contribution
of the candidates from \citeauthor{Ajiki03} to our volume and combine
the two samples, $L^*$ is higher ($\log L^* = 42.66$; case C).
Table~\ref{tab:totlumcomp} summarises the results. Detecting LAEs of
such bright luminosity at this redshift demonstrates the necessity of
wide field surveys, such as WFILAS, to provide a sample of LAEs at the
bright end.

\begin{figure}[!htbp]
  \centering
  \includegraphics[width=0.75\columnwidth, trim=0pt 0pt 0pt 0pt, clip]{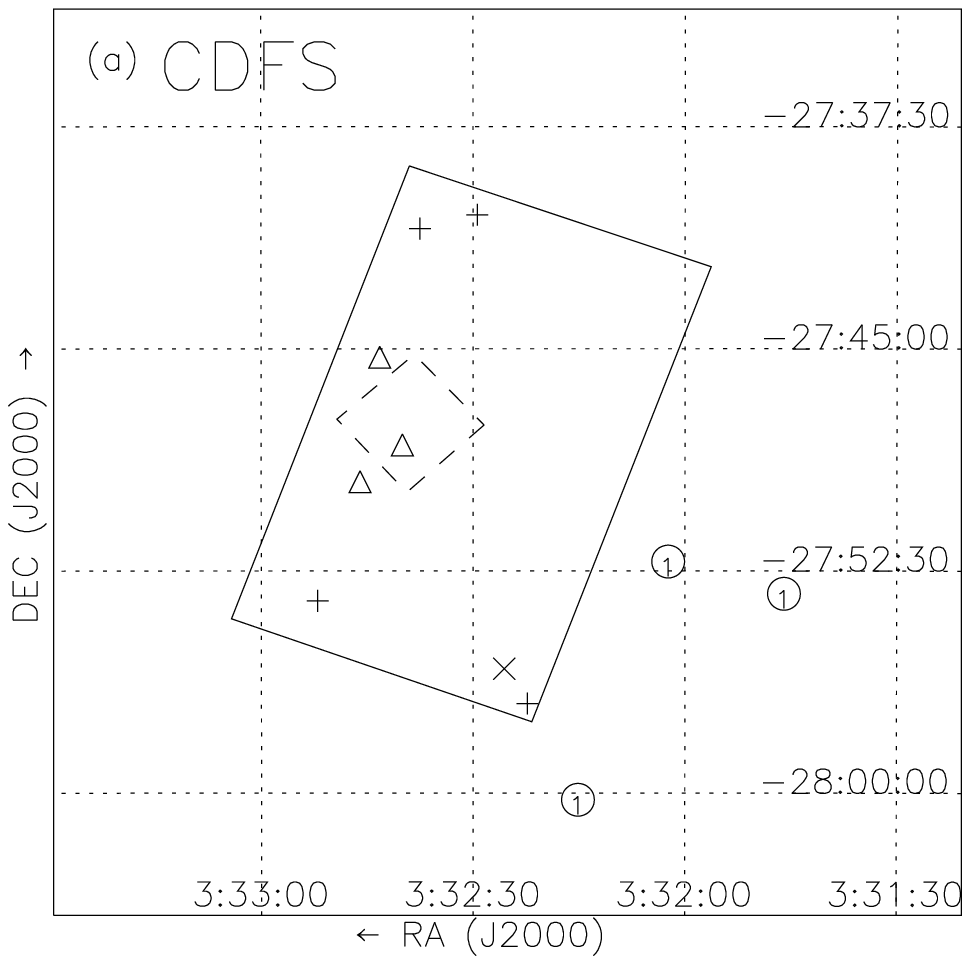}
  \includegraphics[width=0.75\columnwidth, trim=0pt 0pt 0pt 0pt, clip]{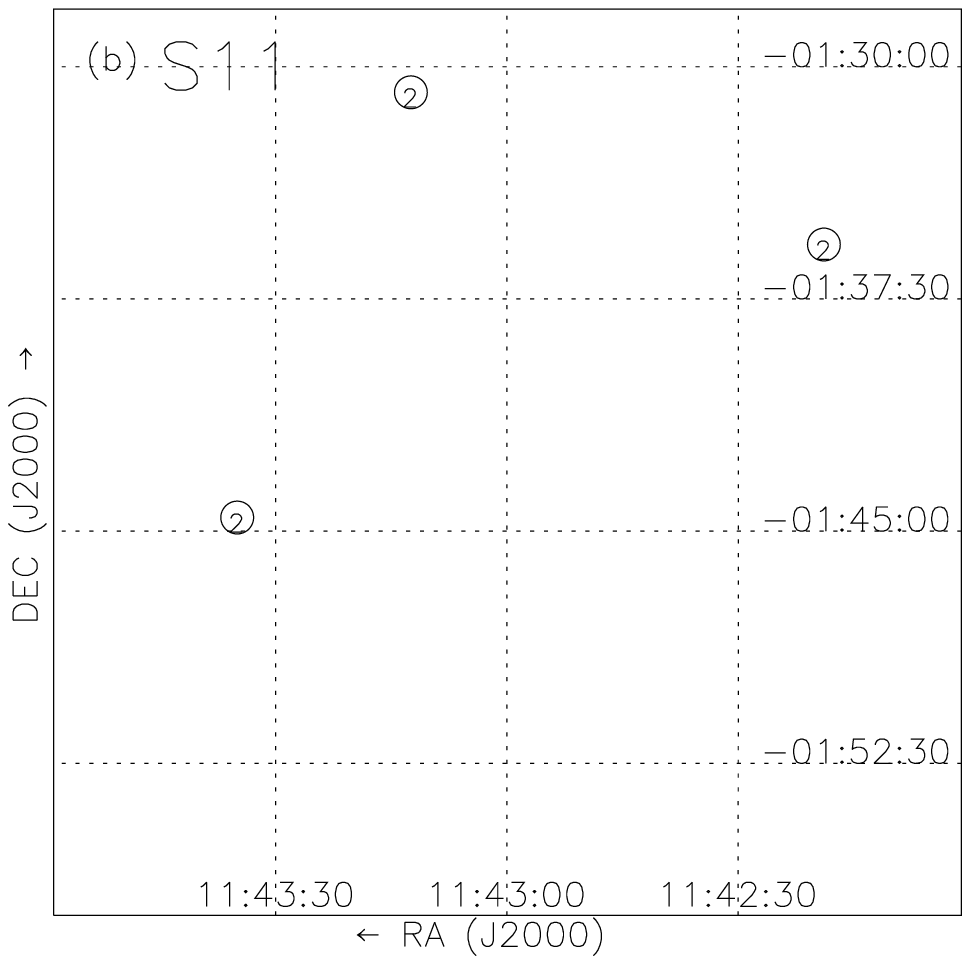}
  \includegraphics[width=0.75\columnwidth, trim=0pt 0pt 0pt 0pt, clip]{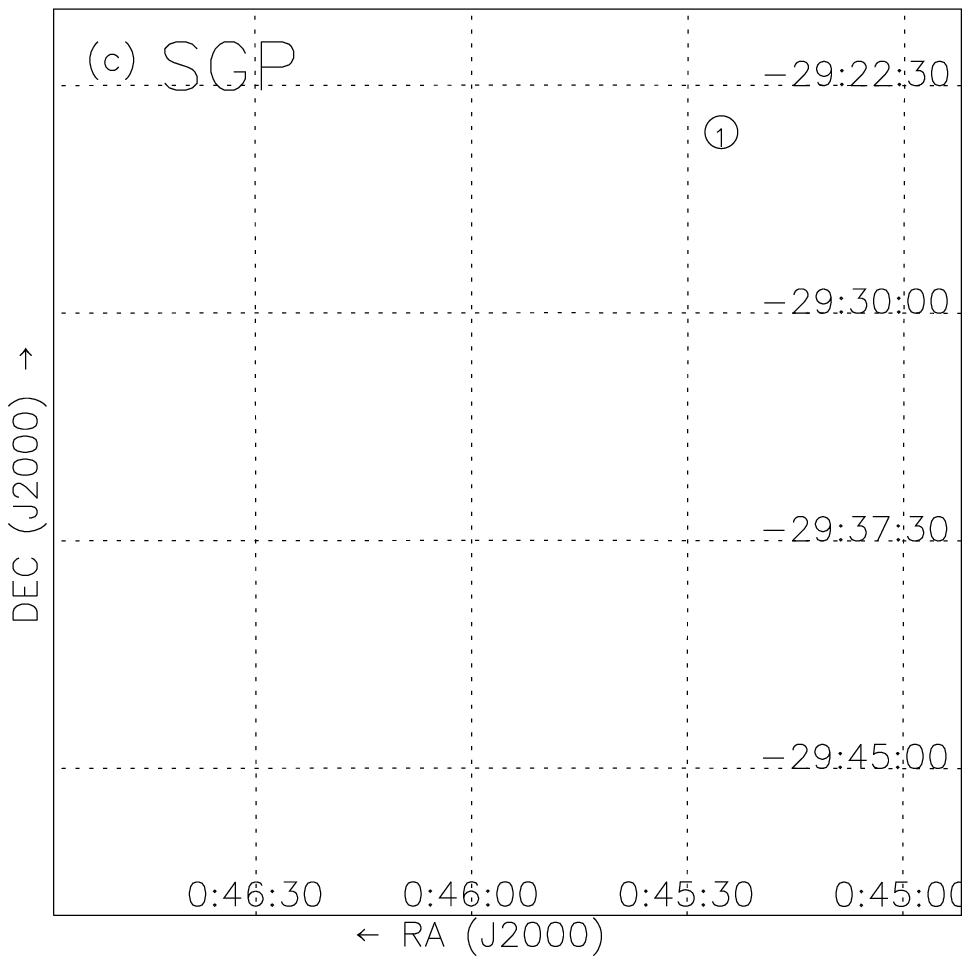}
  \caption{Sky distribution of candidate line emitters per field with
    North up and East to the left for the {\bf a)} CDFS, {\bf b)} S11
    and {\bf c)} SGP fields. The ``1'', ``2'' and ``3'' labels correspond to
    the \na, \nb\ and \nc\ filters, respectively. The gridlines are
    separated by 7\farcm5. In the CDFS field the Hubble Ultra Deep
    Field ({\it dashed}) and GOODS-S ({\it solid}) have been
    indicated. The confirmed $i$-drop galaxy at $z=5.78$ of
    \citet[{\it cross}]{Bunker03}, LAEs of \citet[{\it plus}]{Ajiki05}
    and \citet[{\it triangle}]{Stanway04} are also indicated. In the
    CDFS field there seems to be a overdensity of candidates towards
    the southern part of the field, similar to \citet{Wang05}.}
\label{fig:skydist}
\end{figure}

As a second approach, we tried fitting a Schechter function to the
combined WFILAS and \citet{Ajiki03} dataset, using a minimised
$\chi^2$ fit (Fig.~\ref{fig:complumdist}). We did not use the two
lowest luminosity bins of \citet{Ajiki03} to constrain the fit because
these force the function to decline at the faint end. Instead, we set
the faint end slope to $\alpha$\,=\,$-1.53$, similar to the \ha\
luminosity function at $z\sim0.24$ from \citet{Fujita03}, on which
\citeauthor{Ajiki03} based their work. Figure~\ref{fig:complumdist}b
shows a strong correlation between $L^*$ and $\phi^*$ due to the slow
turn-over at the bright end.

From the fitting there are three results to conclude. Firstly,
incorporating the four completeness-corrected WFILAS galaxies into the
\citet{Ajiki03} galaxies better constrains the bright end of the
luminosity function. Furthermore, it seems that the current generation
of surveys is only just reaching the volume coverage necessary to
discover LAEs with $L>L^*$. The histogram in
Fig.~\ref{fig:complumdist} shows a decreasing number of sources at the
faint end. At face value, this could suggest that the ionising flux of
the less luminous sources may be insufficient to escape the slowly
expanding envelope of neutral hydrogen that surrounds the \Hii\ region
in the LAE. Consequently, the sources are undetected and the faint end
of the luminosity distribution decreases. However, it is difficult to
detect faint LAEs and so the possibility of detection incompleteness
cannot be ruled out.

Figure~\ref{fig:skydist} shows the sky distribution of our candidates
in each field. All candidates but one are in the CDFS and S11 fields.
The only candidate in the SGP field is brighter than the candidates in
the other fields (line flux $\sim$10$^{-16}$\,\lineunits). The reason
for this is that the \med\ filter for the SGP field has a shorter
exposure time and lower signal-to-noise than the other fields.

In the CDFS field we note that our three candidates appear to be
spatially clustered. Additionally, we note that the confirmed $z=5.78$
$i$-drop galaxy of \citet{Bunker03} is at the same redshift as the
WFILAS candidates in this field, just like four candidate LAEs from a
narrowband survey by \citet{Ajiki05}. We did not detect these four
candidates since they are fainter than the detection limits of WFILAS
in this field. \citet{Wang05} have also done a narrowband survey of
the CDFS field. They also find evidence for an overdensity of
$z\sim5.7$ sources in this field. Similarly, \citet{Malhotra05} find
an overdensity at redshift 5.9\,$\pm$\,0.2 in the HUDF.

\section{Confirmed LAEs\protect\footnote{Based on observations made
    with ESO Telescopes at the Paranal Observatories under programmes
    ID 076.A-0553 and 272.A-5029.}}
\label{sec:confirmed}
In \citet{Westra05} we reported the spectroscopic follow-up of one of
the candidates, \object{J114334.98$-$014433.7} (S11\_13368 in that
paper, hereafter S11\_5236\footnote{The object names are derived from
  \sext\ IDs. Refinements to our detection procedures since
  \citet{Westra05} caused a change in the ID and, therefore, in the
  object name}). It was confirmed to be a LAE at $z=5.721$. Here we
present the spectral confirmation of a new candidate,
\object{J004525.38$-$292402.8} (hereafter SGP\_8884), at $z=5.652$. We
also show its pre-imaging and compare its \lya\ profile to S11\_5236.
SGP\_8884 and S11\_5236 are the only two out of the seven candidates
presented in this paper for which we have obtained spectra.

\subsection{Spectral data reduction}

\begin{figure}[tbp]
  \centering
  \includegraphics[width=\columnwidth, trim=1pt 1pt 1pt 1pt]{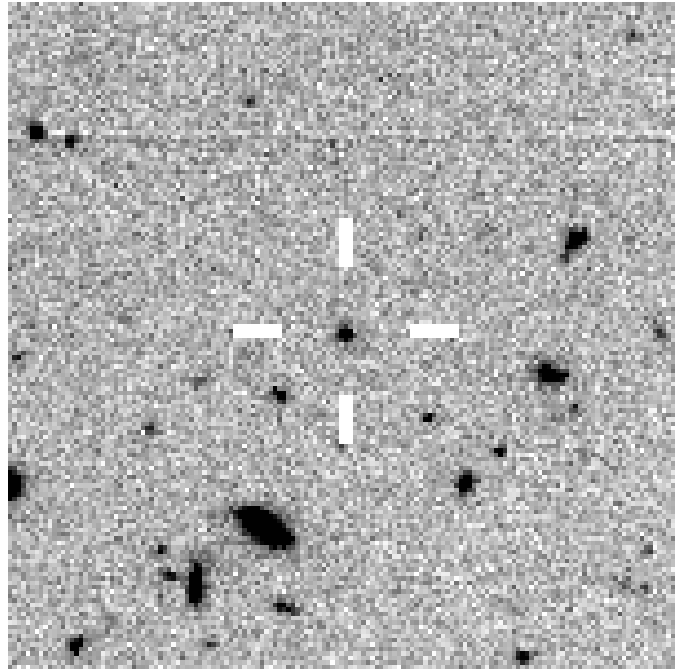}
  \caption{A 38\as\,$\times$\,38\as\ region around the confirmed LAE
    in the SGP field. The image is created from the pre-image taken
    with VLT/FORS2. The image has a pixel scale of
    0\farcs252\,pix$^{-1}$. North is up and East is to the left.}
  \label{fig:SGP_preimage}
\end{figure}

A pre-image with an intermediate band filter ({\it FWHM} = 13\,nm) centred
at 815\,nm was taken with VLT/FORS2 on 2005 August 9. The
0\farcs252\,pix$^{-1}$ plate scale undersamples the $\lesssim$0.5\as
stellar point spread function of the frames which were taken during
excellent seeing. SGP\_8884 is unresolved, implying that the
{\it FWHM} of the emitting region is $\le$2.2\,kpc. A
38\as\,$\times$\,38\as\ region around the object is shown in
Fig.~\ref{fig:SGP_preimage}.

The spectroscopy consists of four exposures of 900\,s, taken on
2005 October 3 with FORS2 using the 1028z grism and a 1\as\ slit. The
frames were overscan subtracted and flatfielded. They were combined by
summing individual frames, thereby removing cosmic rays in the
process.

The spectrum was flux calibrated using a standard star (HD\,49798)
taken with a 5\as\ slit and corrected for slit-loss. This was
calculated assuming a Gaussian source profile with a {\it FWHM} of 0\farcs72
as measured from the spatial direction of the spectrum. The flux lost
due to the 1\as\ slit was calculated and added to the spectrum of the
object.

\subsection{Line fitting}

\begin{figure}[tbp]
  \centering
  \includegraphics[width=\columnwidth, trim=4pt 8pt 19pt 5pt, clip]{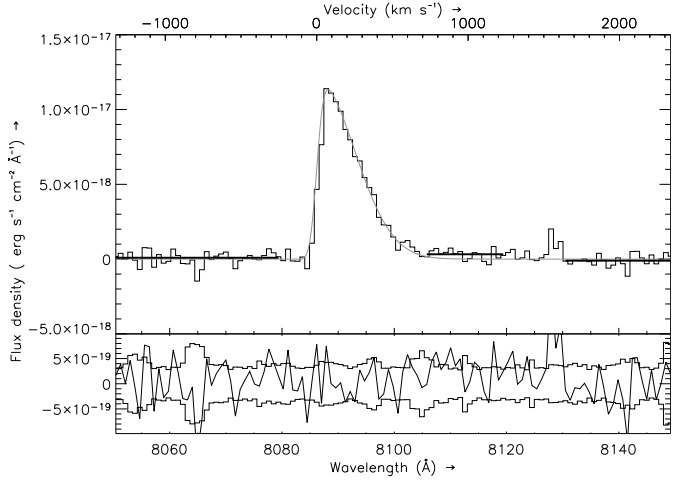}
  \caption{({\it Top}) Flux calibrated spectrum of the confirmed
    candidate LAE SGP\_8884, the brightest candidate in our sample.
    The histogram shows the observed spectrum. Indicated in grey is the
    best-fitting single component model after convolution with the
    instrumental profile. The heavy bold lines indicate three regions
    for which we have calculated a mean continuum. ({\it Bottom})
    Residuals from the observed data minus model fit. The histograms
    indicate the 1$\sigma$-error spectrum from the observed data,
    which includes both sky- and Poisson noise. The feature at
    8125\,\AA\ is due to a remnant cosmic ray from one of the spectral
    frames.}
  \label{fig:fitSGP}
\end{figure}

\begin{figure}[tbp]
  \centering
  \includegraphics[width=\columnwidth, trim=50pt 104pt 5pt 120pt, clip]{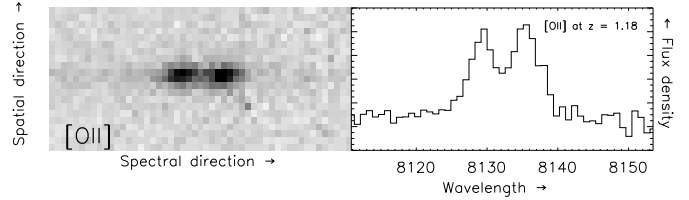}
  \caption{({\it Left}) Two dimensional spectrum of an \oii\ emitting
    galaxy at $z=1.18$ in the SGP field. ({\it Right}) The extracted
    one dimensional spectrum. We are able to easily resolve the \Oii\
    lines with the available resolution. Both spectra are background
    subtracted.}
  \label{fig:OII}
\end{figure}

Figure~\ref{fig:fitSGP} shows the reduced spectrum of SGP\_8884
alongside its best model fit. The spectrum has an asymmetric line
profile, similar to our previously confirmed candidate LAE
\citep{Westra05}. It unlikely originates from a redshifted \oii\ line
at $z\sim1.2$ because the resolution of our spectrum is high enough to
resolve the \Oii. Figure~\ref{fig:OII} shows the spectrum of one such
\oii\ emitter at $z=1.18$ which was included in the same observations
as SGP\_8884. Furthermore, we do not find any other spectral features
in our spectrum, such as \hb\ or \nii, which could classify the
emission coming from a lower redshift galaxy. Hence, we identify the
line as \lya\ at $z=5.652$. With a total spectral line flux of
(1.0\,$\pm$\,0.1)\,$\times$10$^{-16}$\,\lineunits\ (slit-loss
corrected), SGP\_8884 is the brightest LAE at redshift $\sim$5.7 to
date. The line flux derived from the spectrum is consistent with the
flux derived from narrowband photometry
(9.5\,$\pm$\,1.4)\,$\times$\,$\pow{-17}$\,\lineunits, which is given
in Table~\ref{tab:cands}. The spectral line flux corresponds to a line
luminosity of $L_{\rm line}$\,=\,3.5\,$\times$\,$\pow{43}$\,\ergs\ and
a star formation rate of 32\,$M_\odot$\,yr$^{-1}$, using the star
formation conversion rate of \citet{Ajiki03}. If we adopt
$\sim$16\,pixels (=\,32\,kpc$^2$) as an upper limit to the size of
the emitting region, we derive a star formation rate surface density
of $\Sigma_*$\,$\gtrsim$\,1\,$M_\odot$\,yr$^{-1}$\,kpc$^{-2}$.

Following earlier works \citep[e.g.][]{Dawson02,Hu04,Westra05} we
fitted a single component model to the \lya\ line SGP\_8884. The model
consists of a truncated Gaussian with complete absorption blueward of
the \lya\ line centre. We find an excess of flux in the observed data
compared to the model around 8110\,\AA. This suggests the presence of
a second line component redward of the main peak. To test this, we
measured the mean continuum levels, both red- and blueward of the
line, as well as across the red-flanking region of the line. The
continuum is calculated as the weighted mean of the flux density over
this region. This yields for continuum in the red-flanking region a
flux density of (3.2\,$\pm$\,0.8)\,$\times$\,10$^{-19}$\,\ergsPerAng.
Red- and blueward of the \lya\ line the continuum is
(-1.0\,$\pm$\,0.8)\,$\times$\,10$^{-19}$\,\ergsPerAng\ and
(0.9\,$\pm$\,0.6)\,$\times$\,10$^{-19}$\,\ergsPerAng, respectively.
These continuum levels are indicated by the heavy bold lines in
Fig.~\ref{fig:fitSGP}. The lower limit for the rest frame equivalent
width derived from the continuum of the red flank is 46\,\AA. The rest
frame equivalent width derived from the 2$\sigma$ upper limit of the
continuum redward of the line is 125\,\AA.

\begin{figure*}[tbp]
  \centering
  \includegraphics[width=\textwidth, trim=2pt 8pt 15pt 5pt, clip]{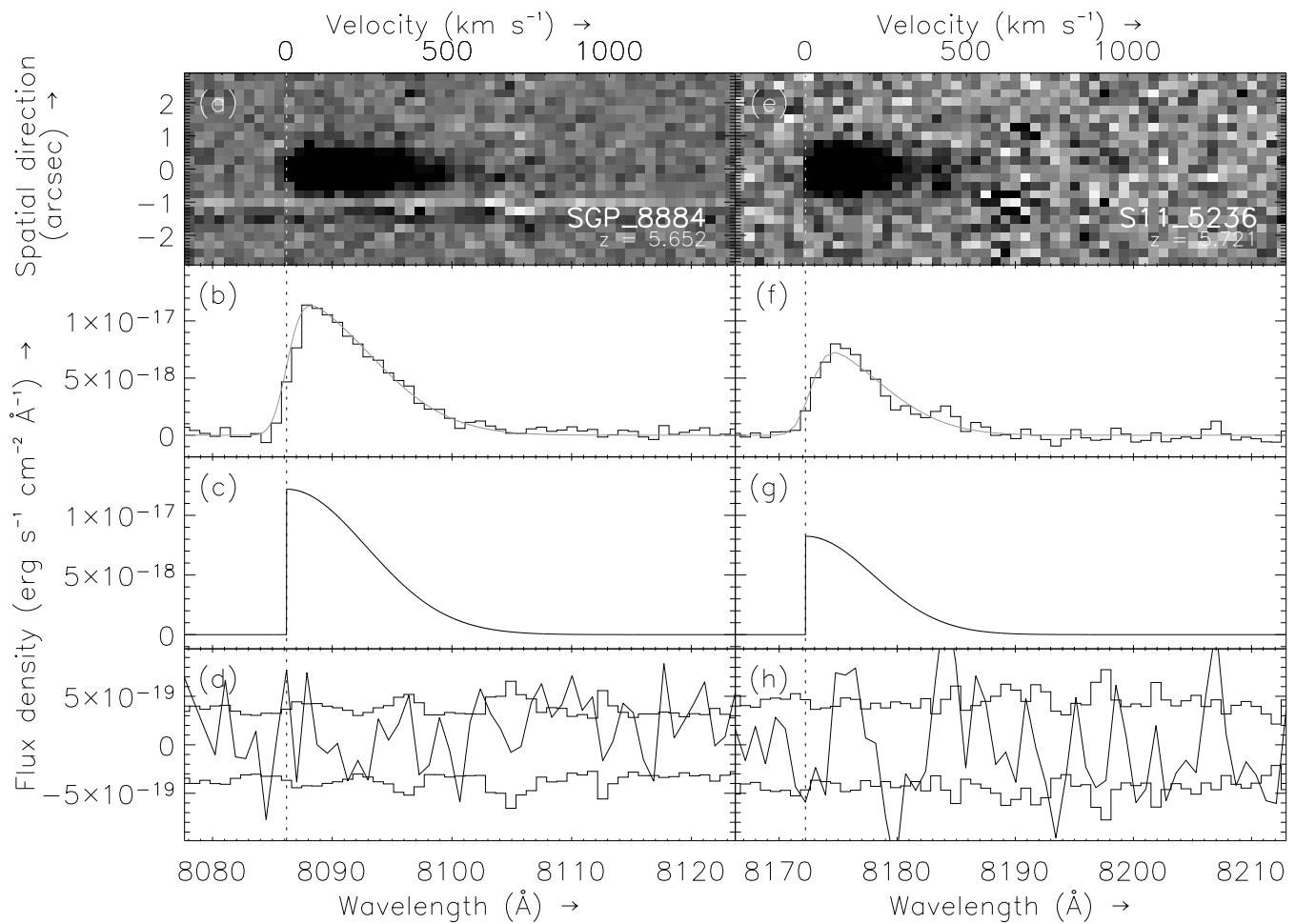}
  \caption{Comparison of the \lya\ line profiles of the two WFILAS
    sources, SGP\_8884 and S11\_5236. {\bf a)} Two dimensional
    background-subtracted spectrum of SGP\_8884. {\bf b)} Observed
    \lya\ line ({\it histograms}) with the best-fitting one component
    model ({\it grey solid line}). {\bf c)} Same model line profile as
    in {\bf b)} but before convolution with the instrument profile.
    {\bf d)} Observed data minus model fit (as plotted in {\bf b)}).
    Also shown ({\it histograms}) is the 1$\sigma$-error spectrum
    from the observed data, which includes both sky- and Poisson
    noise. Panels {\bf e)} through {\bf h)} show the same for
    S11\_5236. The horizontal axes show both the wavelength (in \AA;
    {\it bottom}) and velocity offset from the centre of the full
    Gaussian of the \lya\ line (in \kms; {\it top})}
  \label{fig:fitSGPS11}
\end{figure*}

\begin{table}[tbp]
  \centering
  \newcommand{\thistabwidth}{0.35cm}
  \begin{tabular}{c@{\hspace{\thistabwidth}}c@{\hspace{\thistabwidth}}c@{\hspace{\thistabwidth}}c@{\hspace{\thistabwidth}}c@{\hspace{\thistabwidth}}c@{\hspace{\thistabwidth}}c}
    \hline
    \hline
    Component & $\lambda_c$ & $f_{\rm peak} $ &
    \multicolumn{2}{c}{{\it FWHM}} & \multicolumn{2}{c}{$\Delta v$}\\
    (1) & (2) & (3) & (4) & (5) & (6) & (7)\\
    \hline
    \multicolumn{7}{l}{SGP\_8884 single component}\\
    Single peak & 8086.2 & 1.2\,$\times$\,10$^{-17}$ & 15.7 & 580 & & \\
    \hline
    \multicolumn{7}{l}{S11\_5236 single component}\\
    Single peak & 8172.2 & 8.3\,$\times$\,10$^{-18}$ & 13.5 & 495 & & \\
    \hline
    \multicolumn{7}{l}{S11\_5236 double component, ``broad''}\\
    Main peak & 8173.1 & 8.0\,$\times$\,10$^{-18}$ & 11.3 & 413 & & \\
    Red peak & 8184.1 & 1.9\,$\times$\,10$^{-18}$ & 2.3 & 85 & +11.1 & +406\\
    \hline
    \multicolumn{7}{l}{S11\_5236 double component, ``narrow''}\\
    Main peak & 8173.1 & 8.1\,$\times$\,10$^{-18}$ & 11.2 & 413 & & \\
    Red peak & 8184.1 & 4.8\,$\times$\,10$^{-18}$ & 0.5 & 18 & +11.0 & +403\\
    \hline

  \end{tabular}
  \caption{Parameters for the single component model to SGP\_8884
    before convolution with the instrumental profile. We also include
    the parameters for the single component and the two double
    component models of the previously confirmed LAE S11\_5236
    \citep{Westra05}. These parameters differ slightly from
    \citeauthor{Westra05}, since we have subsequently corrected the
    spectrum of S11\_5236 for slit-losses. \protect \\
    {\it Notes}: (1) component of the fit, (2) central wavelength of
    the fitted component in \AA, (3) peak flux density in \ergsPerAng,
    (4) and (5) {\it FWHM} of full Gaussian of the profile in \AA\ and \kms,
    respectively, (6) and (7) velocity shift of the second component
    in \AA\ and \kms.}
  \label{tab:fitpars}
\end{table}

To see if the excess of flux in the red flank of the \lya\ line can be
explained by an outflow, we fit a second Gaussian component to the
spectrum of SGP\_8884, as we did to the spectrum of S11\_5236 in
\citet{Westra05}. This yields an extremely faint and broad second
component ($f_{\rm peak}$\,$\sim$5\,$\times$\,10$^{-19}$\,\ergsPerAng\
and {\it FWHM}\,$\sim$1700\,\kms). The precise parameters for the red
component are difficult to constrain given its faint and broad
profile. The parameters from the single component model for SGP\_8884
and the single and double component models for S11\_5236 are given in
Table~\ref{tab:fitpars}.

\subsection{Discussion/Comparison}

The \lya\ emission we see is due to intense star formation rates
synonymous with local starburst galaxies. Star formation rates per
unit area in excess of 0.1\,$M_\odot$\,yr$^{-1}$\,kpc$^2$ are prone to
produce large scale outflows of neutral hydrogen from a galaxy,
powered by the supernovae and stellar winds of massive stars
\citep{Heckman02}. The most efficient way for \lya\ to escape from the
compact star forming regions is due to scattering of the photons by
the entrained neutral hydrogen \citep{Chen94}. The kinematics and
orientation of the outflowing neutral hydrogen can alter the \lya\
profile by absorbing photons bluer if along the line of sight, or
backscattering redder than \lya\ if behind and receding
\citep[e.g.][]{Dawson02}. \lya\ emission can also arise when large
scale shocks from starburst winds impinge on clumps ($\sim$100\,pc)
of condensed gas accreting onto the halo \citep{BlandHawthorn04}.

Most examples of asymmetric \lya\ emission at $z\sim6$ show an
extended tail implying backscattering over a fairly wide range of
velocities beyond the central \lya\ emission \citep[e.g. Fig.~9
of][]{Hu04}. The limiting physical size of SGP\_8884
({\it FWHM}\,$<$\,2.2\,kpc) is consistent with the scale of emitting regions
in the local starburst galaxy M82 which span 0.5 to 1\,kpc
\citep{Courvoisier90,Blecha90}. This, and the scale of its outflow,
make it fairly typical of both the starbursting sources seen at
$z\sim6$ and their local counterparts.

The tentative discovery of a second component in S11\_5236
\citep{Westra05} could be explained by either an expanding shell of
neutral hydrogen \citep{Dawson02,Ahn03}, or by infall of the IGM onto
the LAE \citep{Dijkstra05}. The flux of the intrinsic \lya\ line
depends heavily on the model. It is suggested that the total intrinsic
\lya\ flux emerging from these sources is underestimated by an order
of magnitude (e.g. \citeauthor{Dijkstra05}). Therefore, the star
formation rates derived from the observed \lya\ lines could be heavily
underestimated.

Figure~\ref{fig:fitSGPS11} shows a comparison between the line
profiles of the two LAEs discovered with WFILAS. S11\_5236 differs
from SGP\_8884 in that a clear peak, $\sim$20\,$-$\,90\,\kms\ wide,
is seen $\sim$400\,\kms\ redward of \lya\ \citep{Westra05}. The red
component is narrower ($\sim$15\%) and relatively stronger than
SGP\_8884. The difference in the width of the red component is even
more pronounced ($\sim$30\%) when we compare the main peak of the
two-component fits to the spectrum of S11\_5236 to the single peak of
the one-component fits to the spectrum of SGP\_8884. This can clearly
be seen in panels a and e of Fig.~\ref{fig:fitSGPS11}.

Ultimately, such outflows are thought to be responsible for the
chemical enrichment of the IGM by $z\sim6$ \citep{Aguirre01}. Outflows
are a process facilitating the escape of UV photons, which are the
origin for the UV background \citep{Madau99}.

\section{Summary}
\label{sec:summary}
In this paper we have presented the Wide Field Imager Lyman-Alpha
Search (WFILAS), which uses a combination of \mbox{narrow-,}
intermediate and broadband filters on the ESO/MPI 2.2\,m telescope to
search for LAEs at redshift $z\sim5.7$. This search has resulted in
seven bright ($L$\,$\geq$1.1$\times$10$^{43}$\,\ergs) candidate
galaxies across three fields spanning almost 0.8 sq. degree.

Most of our candidates are in the regimes of bright luminosities,
beyond the reach of less voluminous surveys. Adding our candidates to
those of earlier such surveys results in an integrated luminosity
density $\mathcal{L}$ $\sim$30\% higher than found by such surveys
alone. We also find potential clustering in our CDFS field, supporting
overdensities discovered by other surveys. Spectroscopic follow-up for
confirmation in this area will be crucial.

Two candidates have been confirmed to be LAEs at $z\sim5.7$ by means
of spectroscopy. One of these galaxies is the brightest LAEs at this
redshift. The broad, asymmetric profiles of the \lya\ line of both
objects are consistent with neutral hydrogen backscattering of a
central starbursting source.

\begin{acknowledgements}
  The authors wish to thank the Max-Planck-Institut f\"ur Astronomie
  and the DDT grant of the European Southern Observatory for providing
  the narrow band filters which are crucial to the WFILAS survey. The
  broadband and part of the intermediate band data were kindly
  provided by the COMBO-17 team \citep{Wolf04}. We also like to thank
  the anonymous referee for his/her useful suggestions and comments.
  E.W. wishes to thank A.~Frebel for her useful comments and
  discussions regarding this paper and the Astronomical Society of
  Australia Travel Grant. D.H.J. is supported as a Research Associate
  by the Australian Research Council Discovery-Projects Grant
  (DP-0208876), administered by the Australian National University.
  C.W. is supported by a PPARC Advanced Fellowship.
\end{acknowledgements}

\bibliographystyle{aa}
\bibliography{wfilas.bib,astroph.bib}

\begin{thebibliography}{61}
\expandafter\ifx\csname natexlab\endcsname\relax\def\natexlab#1{#1}\fi

\bibitem[{{Adelberger} {et~al.}(2003){Adelberger}, {Steidel}, {Shapley}, \&
  {Pettini}}]{Adelberger03}
{Adelberger}, K.~L., {Steidel}, C.~C., {Shapley}, A.~E., \& {Pettini}, M. 2003,
  \apj, 584, 45

\bibitem[{{Aguirre} {et~al.}(2001){Aguirre}, {Hernquist}, {Schaye}, {Weinberg},
  {Katz}, \& {Gardner}}]{Aguirre01}
{Aguirre}, A., {Hernquist}, L., {Schaye}, J., {et~al.} 2001, \apj, 560, 599

\bibitem[{{Ahn} {et~al.}(2003){Ahn}, {Lee}, \& {Lee}}]{Ahn03}
{Ahn}, S., {Lee}, H., \& {Lee}, H.~M. 2003, \mnras, 340, 863

\bibitem[{{Ajiki} {et~al.}(2005){Ajiki}, {Mobasher}, {Taniguchi}, {Shioya},
  {Nagao}, {Murayama}, \& {Sasaki}}]{Ajiki05}
{Ajiki}, M., {Mobasher}, B., {Taniguchi}, Y., {et~al.} 2005,
  arXiv:astro-ph/0510672

\bibitem[{{Ajiki} {et~al.}(2003){Ajiki}, {Taniguchi}, {Fujita}, {Shioya},
  {Nagao}, {Murayama}, {Yamada}, {Umeda}, \& {Komiyama}}]{Ajiki03}
{Ajiki}, M., {Taniguchi}, Y., {Fujita}, S.~S., {et~al.} 2003, \aj, 126, 2091

\bibitem[{{Allen} {et~al.}(2005){Allen}, {Moustakas}, {Dalton}, {MacDonald},
  {Blake}, {Clewley}, {Heymans}, \& {Wegner}}]{Allen05}
{Allen}, P.~D., {Moustakas}, L.~A., {Dalton}, G., {et~al.} 2005, \mnras, 360,
  1244

\bibitem[{{Baade} {et~al.}(1999){Baade}, {Meisenheimer}, {Iwert}, {Alonso},
  {Augusteijn}, {Beletic}, {Bellemann}, {Benesch}, {Boehm}, {Boehnhardt},
  {Brewer}, {Deiries}, {Delabre}, {Donaldson}, {Dupuy}, {Franke}, {Gerdes},
  {Gilliotte}, {Grimm}, {Haddad}, {Hess}, {Ihle}, {Klein}, {Lenzen}, {Lizon},
  {Mancini}, {Muench}, {Pizarro}, {Prado}, {Rahmer}, {Reyes}, {Richardson},
  {Robledo}, {Sanchez}, {Silber}, {Sinclaire}, {Wackermann}, \&
  {Zaggia}}]{Baade99}
{Baade}, D., {Meisenheimer}, K., {Iwert}, O., {et~al.} 1999, The Messenger, 95,
  15

\bibitem[{{Barger} {et~al.}(2003){Barger}, {Cowie}, {Capak}, {Alexander},
  {Bauer}, {Brandt}, {Garmire}, \& {Hornschemeier}}]{Barger03}
{Barger}, A.~J., {Cowie}, L.~L., {Capak}, P., {et~al.} 2003, \apjl, 584, L61

\bibitem[{{Becker} {et~al.}(2001){Becker}, {Fan}, {White}, {Strauss},
  {Narayanan}, {Lupton}, {Gunn}, {Annis}, {Bahcall}, {Brinkmann}, {Connolly},
  {Csabai}, {Czarapata}, {Doi}, {Heckman}, {Hennessy}, {Ivezi{\' c}}, {Knapp},
  {Lamb}, {McKay}, {Munn}, {Nash}, {Nichol}, {Pier}, {Richards}, {Schneider},
  {Stoughton}, {Szalay}, {Thakar}, \& {York}}]{Becker01}
{Becker}, R.~H., {Fan}, X., {White}, R.~L., {et~al.} 2001, \aj, 122, 2850

\bibitem[{{Bertin} \& {Arnouts}(1996)}]{Bertin96}
{Bertin}, E. \& {Arnouts}, S. 1996, \aaps, 117, 393

\bibitem[{{Bessell}(1999)}]{Bessell99}
{Bessell}, M.~S. 1999, \pasp, 111, 1426

\bibitem[{{Bland-Hawthorn} \& {Nulsen}(2004)}]{BlandHawthorn04}
{Bland-Hawthorn}, J. \& {Nulsen}, P.~E.~J. 2004, arXiv:astro-ph/0404241

\bibitem[{{Blecha} {et~al.}(1990){Blecha}, {Golay}, {Huguenin}, {Reichen}, \&
  {Bersier}}]{Blecha90}
{Blecha}, A., {Golay}, M., {Huguenin}, D., {Reichen}, D., \& {Bersier}, D.
  1990, \aap, 233, L9

\bibitem[{{Bouwens} {et~al.}(2005){Bouwens}, {Illingworth}, {Blakeslee}, \&
  {Franx}}]{Bouwens05}
{Bouwens}, R.~J., {Illingworth}, G.~D., {Blakeslee}, J.~P., \& {Franx}, M.
  2005, arXiv:astro-ph/0509641

\bibitem[{{Bunker} {et~al.}(2004){Bunker}, {Stanway}, {Ellis}, \&
  {McMahon}}]{Bunker04}
{Bunker}, A.~J., {Stanway}, E.~R., {Ellis}, R.~S., \& {McMahon}, R.~G. 2004,
  \mnras, 355, 374

\bibitem[{{Bunker} {et~al.}(2003){Bunker}, {Stanway}, {Ellis}, {McMahon}, \&
  {McCarthy}}]{Bunker03}
{Bunker}, A.~J., {Stanway}, E.~R., {Ellis}, R.~S., {McMahon}, R.~G., \&
  {McCarthy}, P.~J. 2003, \mnras, 342, L47

\bibitem[{{Chen} \& {Neufeld}(1994)}]{Chen94}
{Chen}, W.~L. \& {Neufeld}, D.~A. 1994, \apj, 432, 567

\bibitem[{{Courvoisier} {et~al.}(1990){Courvoisier}, {Reichen}, {Blecha},
  {Golay}, \& {Huguenin}}]{Courvoisier90}
{Courvoisier}, T.~J.-L., {Reichen}, M., {Blecha}, A., {Golay}, M., \&
  {Huguenin}, D. 1990, \aap, 238, 63

\bibitem[{{Cuby} {et~al.}(2003){Cuby}, {Le F{\` e}vre}, {McCracken},
  {Cuillandre}, {Magnier}, \& {Meneux}}]{Cuby03}
{Cuby}, J.-G., {Le F{\` e}vre}, O., {McCracken}, H., {et~al.} 2003, \aap, 405,
  L19

\bibitem[{{Dawson} {et~al.}(2004){Dawson}, {Rhoads}, {Malhotra}, {Stern},
  {Dey}, {Spinrad}, {Jannuzi}, {Wang}, \& {Landes}}]{Dawson04}
{Dawson}, S., {Rhoads}, J.~E., {Malhotra}, S., {et~al.} 2004, \apj, 617, 707

\bibitem[{{Dawson} {et~al.}(2002){Dawson}, {Spinrad}, {Stern}, {Dey}, {van
  Breugel}, {de Vries}, \& {Reuland}}]{Dawson02}
{Dawson}, S., {Spinrad}, H., {Stern}, D., {et~al.} 2002, \apj, 570, 92

\bibitem[{{Dijkstra} {et~al.}(2005){Dijkstra}, {Haiman}, \&
  {Spaans}}]{Dijkstra05}
{Dijkstra}, M., {Haiman}, Z., \& {Spaans}, M. 2005, arXiv:astro-ph/0510409

\bibitem[{{Djorgovski} {et~al.}(2001){Djorgovski}, {Castro}, {Stern}, \&
  {Mahabal}}]{Djorgovski01}
{Djorgovski}, S.~G., {Castro}, S., {Stern}, D., \& {Mahabal}, A.~A. 2001,
  \apjl, 560, L5

\bibitem[{{Fan} {et~al.}(2002){Fan}, {Narayanan}, {Strauss}, {White}, {Becker},
  {Pentericci}, \& {Rix}}]{Fan02}
{Fan}, X., {Narayanan}, V.~K., {Strauss}, M.~A., {et~al.} 2002, \aj, 123, 1247

\bibitem[{{Feldmeier} {et~al.}(2002){Feldmeier}, {Mihos}, {Morrison}, {Rodney},
  \& {Harding}}]{Feldmeier02}
{Feldmeier}, J.~J., {Mihos}, J.~C., {Morrison}, H.~L., {Rodney}, S.~A., \&
  {Harding}, P. 2002, \apj, 575, 779

\bibitem[{{Foucaud} {et~al.}(2003){Foucaud}, {McCracken}, {Le F{\`e}vre},
  {Arnouts}, {Brodwin}, {Lilly}, {Crampton}, \& {Mellier}}]{Foucaud03}
{Foucaud}, S., {McCracken}, H.~J., {Le F{\`e}vre}, O., {et~al.} 2003, \aap,
  409, 835

\bibitem[{{Fujita} {et~al.}(2003){Fujita}, {Ajiki}, {Shioya}, {Nagao},
  {Murayama}, {Taniguchi}, {Umeda}, {Yamada}, {Yagi}, {Okamura}, \&
  {Komiyama}}]{Fujita03}
{Fujita}, S.~S., {Ajiki}, M., {Shioya}, Y., {et~al.} 2003, \apjl, 586, L115

\bibitem[{{Giavalisco} \& {Dickinson}(2001)}]{Giavalisco01}
{Giavalisco}, M. \& {Dickinson}, M. 2001, \apj, 550, 177

\bibitem[{{Gnedin} \& {Ostriker}(1997)}]{Gnedin97}
{Gnedin}, N.~Y. \& {Ostriker}, J.~P. 1997, \apj, 486, 581

\bibitem[{{Haiman} \& {Loeb}(1998)}]{Haiman98}
{Haiman}, Z. \& {Loeb}, A. 1998, \apj, 503, 505

\bibitem[{{Heckman}(2002)}]{Heckman02}
{Heckman}, T.~M. 2002, in ASP Conf. Ser. 254: Extragalactic Gas at Low
  Redshift, 292--+

\bibitem[{{Hildebrandt} {et~al.}(2005){Hildebrandt}, {Bomans}, {Erben},
  {Schneider}, {Schirmer}, {Czoske}, {Dietrich}, {Schrabback}, {Simon},
  {Dettmar}, {Haberzettl}, {Hetterscheidt}, \& {Cordes}}]{Hildebrandt05}
{Hildebrandt}, H., {Bomans}, D.~J., {Erben}, T., {et~al.} 2005, \aap, 441, 905

\bibitem[{{Hippelein} {et~al.}(2003){Hippelein}, {Maier}, {Meisenheimer},
  {Wolf}, {Fried}, {von Kuhlmann}, {K{\" u}mmel}, {Phleps}, \& {R{\"
  o}ser}}]{Hippelein03}
{Hippelein}, H., {Maier}, C., {Meisenheimer}, K., {et~al.} 2003, \aap, 402, 65

\bibitem[{{Hu} {et~al.}(2004){Hu}, {Cowie}, {Capak}, {McMahon}, {Hayashino}, \&
  {Komiyama}}]{Hu04}
{Hu}, E.~M., {Cowie}, L.~L., {Capak}, P., {et~al.} 2004, \aj, 127, 563

\bibitem[{{Hu} {et~al.}(1998){Hu}, {Cowie}, \& {McMahon}}]{Hu98}
{Hu}, E.~M., {Cowie}, L.~L., \& {McMahon}, R.~G. 1998, \apjl, 502, L99+

\bibitem[{{Kodaira} {et~al.}(2003){Kodaira}, {Taniguchi}, {Kashikawa}, {Kaifu},
  {Ando}, {Karoji}, {Ajiki}, {Akiyama}, {Aoki}, {Doi}, {Fujita}, {Furusawa},
  {Hayashino}, {Imanishi}, {Iwamuro}, {Iye}, {Kawabata}, {Kobayashi}, {Kodama},
  {Komiyama}, {Kosugi}, {Matsuda}, {Miyazaki}, {Mizumoto}, {Motohara},
  {Murayama}, {Nagao}, {Nariai}, {Ohta}, {Ohyama}, {Okamura}, {Ouchi},
  {Sasaki}, {Sekiguchi}, {Shimasaku}, {Shioya}, {Takata}, {Tamura}, {Terada},
  {Umemura}, {Usuda}, {Yagi}, {Yamada}, {Yasuda}, \& {Yoshida}}]{Kodaira03}
{Kodaira}, K., {Taniguchi}, Y., {Kashikawa}, N., {et~al.} 2003, \pasj, 55, L17

\bibitem[{{Labb{\' e}} {et~al.}(2003){Labb{\' e}}, {Franx}, {Rudnick},
  {Schreiber}, {Rix}, {Moorwood}, {van Dokkum}, {van der Werf}, {R{\"
  o}ttgering}, {van Starkenburg}, {van de Wel}, {Kuijken}, \&
  {Daddi}}]{Labbe03}
{Labb{\' e}}, I., {Franx}, M., {Rudnick}, G., {et~al.} 2003, \aj, 125, 1107

\bibitem[{{Madau} {et~al.}(1999){Madau}, {Haardt}, \& {Rees}}]{Madau99}
{Madau}, P., {Haardt}, F., \& {Rees}, M.~J. 1999, \apj, 514, 648

\bibitem[{{Maier} {et~al.}(2003){Maier}, {Meisenheimer}, {Thommes},
  {Hippelein}, {R{\" o}ser}, {Fried}, {von Kuhlmann}, {Phleps}, \&
  {Wolf}}]{Maier03}
{Maier}, C., {Meisenheimer}, K., {Thommes}, E., {et~al.} 2003, \aap, 402, 79

\bibitem[{{Malhotra} {et~al.}(2005){Malhotra}, {Rhoads}, {Pirzkal}, {Haiman},
  {Xu}, {Daddi}, {Yan}, {Bergeron}, {Wang}, {Ferguson}, {Gronwall},
  {Koekemoer}, {Kuemmel}, {Moustakas}, {Panagia}, {Pasquali}, {Stiavelli},
  {Walsh}, {Windhorst}, \& {di Serego Alighieri}}]{Malhotra05}
{Malhotra}, S., {Rhoads}, J.~E., {Pirzkal}, N., {et~al.} 2005, \apj, 626, 666

\bibitem[{{M{\o}ller} \& {Fynbo}(2001)}]{Moeller01}
{M{\o}ller}, P. \& {Fynbo}, J.~U. 2001, \aap, 372, L57

\bibitem[{{Oke} \& {Gunn}(1983)}]{Oke83}
{Oke}, J.~B. \& {Gunn}, J.~E. 1983, \apj, 266, 713

\bibitem[{{Ouchi} {et~al.}(2005){Ouchi}, {Shimasaku}, {Akiyama}, {Sekiguchi},
  {Furusawa}, {Okamura}, {Kashikawa}, {Iye}, {Kodama}, {Saito}, {Sasaki},
  {Simpson}, {Takata}, {Yamada}, {Yamanoi}, {Yoshida}, \& {Yoshida}}]{Ouchi05}
{Ouchi}, M., {Shimasaku}, K., {Akiyama}, M., {et~al.} 2005, \apjl, 620, L1

\bibitem[{{Ouchi} {et~al.}(2004){Ouchi}, {Shimasaku}, {Okamura}, {Furusawa},
  {Kashikawa}, {Ota}, {Doi}, {Hamabe}, {Kimura}, {Komiyama}, {Miyazaki},
  {Miyazaki}, {Nakata}, {Sekiguchi}, {Yagi}, \& {Yasuda}}]{Ouchi04}
{Ouchi}, M., {Shimasaku}, K., {Okamura}, S., {et~al.} 2004, \apj, 611, 685

\bibitem[{{Rhoads} {et~al.}(2003){Rhoads}, {Dey}, {Malhotra}, {Stern},
  {Spinrad}, {Jannuzi}, {Dawson}, {Brown}, \& {Landes}}]{Rhoads03}
{Rhoads}, J.~E., {Dey}, A., {Malhotra}, S., {et~al.} 2003, \aj, 125, 1006

\bibitem[{{Rhoads} \& {Malhotra}(2001)}]{Rhoads01}
{Rhoads}, J.~E. \& {Malhotra}, S. 2001, \apjl, 563, L5

\bibitem[{{Santos} {et~al.}(2004){Santos}, {Ellis}, {Kneib}, {Richard}, \&
  {Kuijken}}]{Santos04}
{Santos}, M.~R., {Ellis}, R.~S., {Kneib}, J., {Richard}, J., \& {Kuijken}, K.
  2004, \apj, 606, 683

\bibitem[{{Schechter}(1976)}]{Schechter76}
{Schechter}, P. 1976, \apj, 203, 297

\bibitem[{{Schlegel} {et~al.}(1998){Schlegel}, {Finkbeiner}, \&
  {Davis}}]{Schlegel98}
{Schlegel}, D.~J., {Finkbeiner}, D.~P., \& {Davis}, M. 1998, \apj, 500, 525

\bibitem[{{Spergel} {et~al.}(2006){Spergel}, {Bean}, {Dore'}, {Nolta},
  {Bennett}, {Hinshaw}, {Jarosik}, {Komatsu}, {Page}, {Peiris}, {Verde},
  {Barnes}, {Halpern}, {Hill}, {Kogut}, {Limon}, {Meyer}, {Odegard}, {Tucker},
  {Weiland}, {Wollack}, \& {Wright}}]{Spergel06}
{Spergel}, D.~N., {Bean}, R., {Dore'}, O., {et~al.} 2006,
  arXiv:astro-ph/0603449

\bibitem[{{Stanway} {et~al.}(2004){Stanway}, {Glazebrook}, {Bunker}, {Abraham},
  {Hook}, {Rhoads}, {McCarthy}, {Boyle}, {Colless}, {Crampton}, {Couch},
  {J{\o}rgensen}, {Malhotra}, {Murowinski}, {Roth}, {Savaglio}, \&
  {Tsvetanov}}]{Stanway04}
{Stanway}, E.~R., {Glazebrook}, K., {Bunker}, A.~J., {et~al.} 2004, \apjl, 604,
  L13

\bibitem[{{Steidel} {et~al.}(1999){Steidel}, {Adelberger}, {Giavalisco},
  {Dickinson}, \& {Pettini}}]{Steidel99}
{Steidel}, C.~C., {Adelberger}, K.~L., {Giavalisco}, M., {Dickinson}, M., \&
  {Pettini}, M. 1999, \apj, 519, 1

\bibitem[{{Steidel} {et~al.}(2000){Steidel}, {Adelberger}, {Shapley},
  {Pettini}, {Dickinson}, \& {Giavalisco}}]{Steidel00}
{Steidel}, C.~C., {Adelberger}, K.~L., {Shapley}, A.~E., {et~al.} 2000, \apj,
  532, 170

\bibitem[{{Stiavelli} {et~al.}(2001){Stiavelli}, {Scarlata}, {Panagia}, {Treu},
  {Bertin}, \& {Bertola}}]{Stiavelli01}
{Stiavelli}, M., {Scarlata}, C., {Panagia}, N., {et~al.} 2001, \apjl, 561, L37

\bibitem[{{Venemans} {et~al.}(2002){Venemans}, {Kurk}, {Miley},
  {R{\"o}ttgering}, {van Breugel}, {Carilli}, {De Breuck}, {Ford}, {Heckman},
  {McCarthy}, \& {Pentericci}}]{Venemans02}
{Venemans}, B.~P., {Kurk}, J.~D., {Miley}, G.~K., {et~al.} 2002, \apjl, 569,
  L11

\bibitem[{{Wang} {et~al.}(2005){Wang}, {Malhotra}, \& {Rhoads}}]{Wang05}
{Wang}, J.~X., {Malhotra}, S., \& {Rhoads}, J.~E. 2005, \apjl, 622, L77

\bibitem[{{Westra} {et~al.}(2005){Westra}, {Jones}, {Lidman}, {Athreya},
  {Meisenheimer}, {Wolf}, {Szeifert}, {Pompei}, \& {Vanzi}}]{Westra05}
{Westra}, E., {Jones}, D.~H., {Lidman}, C.~E., {et~al.} 2005, \aap, 430, L21

\bibitem[{{Williams} {et~al.}(1996){Williams}, {Blacker}, {Dickinson}, {Dixon},
  {Ferguson}, {Fruchter}, {Giavalisco}, {Gilliland}, {Heyer}, {Katsanis},
  {Levay}, {Lucas}, {McElroy}, {Petro}, {Postman}, {Adorf}, \&
  {Hook}}]{Williams96}
{Williams}, R.~E., {Blacker}, B., {Dickinson}, M., {et~al.} 1996, \aj, 112,
  1335

\bibitem[{{Wolf} {et~al.}(2004){Wolf}, {Meisenheimer}, {Kleinheinrich},
  {Borch}, {Dye}, {Gray}, {Wisotzki}, {Bell}, {Rix}, {Cimatti}, {Hasinger}, \&
  {Szokoly}}]{Wolf04}
{Wolf}, C., {Meisenheimer}, K., {Kleinheinrich}, M., {et~al.} 2004, \aap, 421,
  913

\bibitem[{{Yan} \& {Windhorst}(2004)}]{Yan04}
{Yan}, H. \& {Windhorst}, R.~A. 2004, \apjl, 600, L1

\bibitem[{{Zacharias} {et~al.}(2004){Zacharias}, {Urban}, {Zacharias},
  {Wycoff}, {Hall}, {Monet}, \& {Rafferty}}]{Zacharias04}
{Zacharias}, N., {Urban}, S.~E., {Zacharias}, M.~I., {et~al.} 2004, \aj, 127,
  3043

\end{thebibliography}

\begin{landscape}
\begin{table}[!htbp]
\begin{minipage}[htb]{\textheight}
  \renewcommand{\thefootnote}{{\it\alph{footnote}}}
\begin{center}
\begin{tabular}{rlcccccccc}
\hline\hline
\sext\ ID& Object ID                   & $B$                 & $R$                 & \med\            & \na\                & \nb\                & \nc\                & Line flux                & Luminosity  \\
         &                             &                     &                     &                  &                     &                     &                     & (10$^{-17}$\,\lineunits) & (10$^{43}$\,\ergs) \\
\hline
CDFS\_1864\footnotemark[1] & J033215.14-280013.9 & $>$26.25        & $>$26.56        & 24.72 $\pm$ 0.46 & 23.14 $\pm$ 0.26 & $>$24.27        & $>$23.93        &  6.5 $\pm$ 1.5 & 2.3 $\pm$ 0.5\\
CDFS\_4928\footnotemark[1] & J033145.97-275316.4 & $>$26.25        & $>$26.56        & 24.59 $\pm$ 0.41 & 23.38 $\pm$ 0.32 & 24.11 $\pm$ 0.47 & 23.61 $\pm$ 0.41 &  5.2 $\pm$ 1.6 & 1.8 $\pm$ 0.5\\
CDFS\_5388\footnotemark[1] & J033202.37-275211.3 & $>$26.25        & $>$26.56        & 24.70 $\pm$ 0.45 & 23.32 $\pm$ 0.31 & $>$24.27        & $>$23.93        &  5.5 $\pm$ 1.5 & 1.9 $\pm$ 0.5\\
 S11\_5236\footnotemark[1]\footnotemark[2] & J114334.98-014433.7 & $>$26.63        & $>$26.59        & 24.31 $\pm$ 0.42 & $>$24.13        & 23.05 $\pm$ 0.18 & $>$23.74        &  7.0 $\pm$ 1.2 & 2.5 $\pm$ 0.4\\
 S11\_8921\footnotemark[3] & J114218.90-013544.6 & $>$26.63        & 26.38 $\pm$ 0.45 & 23.88 $\pm$ 0.28 & 23.98 $\pm$ 0.47 & 23.41 $\pm$ 0.26 & $>$23.75        &  5.0 $\pm$ 1.2 & 1.8 $\pm$ 0.4\\
S11\_10595 & J114312.46-013049.6 & $>$26.63        & $>$26.60        & 24.44 $\pm$ 0.47 & $>$24.13        & 23.52 $\pm$ 0.28 & $>$23.75        &  4.5 $\pm$ 1.2 & 1.6 $\pm$ 0.4\\
 SGP\_8884\footnotemark[4] & J004525.38-292402.8 & $>$26.07        & $>$26.41        & 23.33 $\pm$ 0.20 & 22.73 $\pm$ 0.16 &                  & $>$24.06        &  9.5 $\pm$ 1.4 & 3.3 $\pm$ 0.5\\

\hline
\end{tabular}
\end{center}
\caption{The candidate list of the WFILAS survey after the selection
  as described in Sect.~\ref{sec:selection}. From left to right are
  the object name, the $B$, $R$, \med, \na, \nb\ and \nc\
  \mbox{{\it AB}-magnitudes}, line flux calculated from the narrowband
  magnitude in which the object was detected and line luminosity. For
  all measurements less than $2\sigma$ the $2\sigma$ upper limit has
  been given.}
\label{tab:cands}

\footnotetext[1]{Galaxy is in the complete sample}

\footnotetext[2]{Confirmed LAE at $z$ = 5.721. See text and
  \citet{Westra05} for details.}

\footnotetext[3]{Signal-to-noise in the range $2-3\sigma$ for $R$ band
  in the 10 pixel aperture, but $<2\sigma$ in the 6 pixel aperture}

\footnotetext[4]{Confirmed LAE at $z$ = 5.652. See text for details.}
\end{minipage} 
\end{table} 

\begin{table}[!htbp]
\begin{center}
\begin{tabular}{lccccccl}
\hline\hline
       & $\alpha$ & $\log \phi^*$ & $\log L^*$  & $\log L_{lim}$ & log $V$ & $\log \mathcal{L}$ & Comment\\
       &          & Mpc$^{-3}$    & \ergs       & \ergs          & Mpc$^3$ & \lumDens           &        \\
\hline
Case A & --       & --            & --          & 42.85          & 5.26     &  39.04       & Sum of the candidates from \citet{Ajiki03}\\
       & -1.53    & -2.62         &  42.61 & 42.85          & 5.26     &  39.04       & Integrated luminosity function down to \citet{Ajiki03} survey limit (7.0$\times$10$^{42}$\,\ergs)\\
       & -1.53    & -2.62         &  42.61 & --             & --       &  40.27       & Integration of the entire luminosity function\\
\\
Case B & --       & --            & --          & 43.26          & 5.71     &  38.36       & Sum of the candidates from completeness corrected WFILAS sample\\
       & -1.53    & -2.62         &  42.74 & 43.26          & 5.71     &  38.36       & Integrated luminosity function down to the limit of the completeness corrected sample (1.8$\times$10$^{43}$\,\ergs)\\
       & -1.53    & -2.62         &  42.74 & --             & --       &  40.39       & Integration of the entire luminosity function\\
\\
Case C & --       & --            & --          & 42.85          & 5.84     &  39.19       & Sum of the combined WFILAS and \citet{Ajiki03} samples low luminosity corrections\\
       & -1.53    & -2.62         &  42.66 & 42.85          & 5.84     &  39.19       & Integrated luminosity function down to \citet{Ajiki03} survey limit (7.0$\times$10$^{42}$\,\ergs)\\
       & -1.53    & -2.62         &  42.66 & --             & --       &  40.32       & Integration of the entire luminosity function\\

\hline
\end{tabular}
\caption{Calculation of the Schechter function parameter $L^*$ and
  luminosity density $\mathcal{L}$ according to \citet{Ajiki03} for
  their sample, our complete sample and the combination of the two.
  For each sample the luminosity density has been derived from the sum
  of the candidate luminosities divided by the corresponding survey
  volume. Then Eq.~(\ref{eq:lumdens}) was solved for $L^*$, with given
  $\alpha$ and $\phi^*$ from \citet{Ajiki03}. Finally, the entire
  luminosity function was integrated to give the final luminosity
  density.}
\label{tab:totlumcomp}
\end{center}
\end{table}
\end{landscape}

\end{document}